\documentclass[journal]{IEEEtran}

\IEEEoverridecommandlockouts                              

\usepackage{multicol}
\usepackage{graphicx}
\usepackage{etoolbox}
\usepackage{amsmath} 
\usepackage{cite}
\usepackage[table,x11names]{xcolor}
\usepackage{soul,color}
\soulregister\cite7
\soulregister\ref7
\soulregister\pageref7
\usepackage{subcaption}
\usepackage[font=footnotesize]{caption}
\usepackage{epstopdf}
\DeclareGraphicsExtensions{.eps}

\begin{document}
\title{Multivariate Multiscale Dispersion Entropy of Biomedical Times Series}

\author{Hamed Azami$^{1,*}$, \textit{Student Member}, \textit{IEEE}, Alberto Fern{\'a}ndez$^{2}$, and Javier Escudero$^{1}$, \textit{Member}, \textit{IEEE}
	\thanks{$^{1}$ H. Azami and J. Escudero are with the Institute for Digital Communications, School of Engineering, The University of Edinburgh, Edinburgh, King's Buildings, EH9 3FB, United Kingdom. (Phone: +44 131 650 5599, emails: hamed.azami@ed.ac.uk, javier.escudero@ed.ac.uk). *Corresponding author.}
		\thanks{$^{2}$ A. Fern{\'a}ndez is with the Departamento de Psiquiatr{\'i}a y Psicolog{\'i}a M{\'e}dica, Universidad Complutense de Madrid, Madrid, Spain. He is also with Laboratorio de Neurociencia Cognitiva y Computacional, Centro de Tecnolog{\'i}a Biom{\'e}dica, Universidad Politecnica de Madrid and Universidad Complutense de Madrid, Madrid, Spain and with the Insitituto de Investigaci{\'o}n Sanitaria San Carlos (IdSSC).}
}%

\maketitle

\begin{abstract}
\textit{Objective}: Due to the non-linearity of numerous biomedical signals, non-linear analysis of multi-channel time series, notably multivariate multiscale entropy (mvMSE), has been extensively used in biomedical signal processing. However, mvMSE has three drawbacks: 1) mvMSE values are either undefined or unreliable for short signals; 2) mvMSE is not fast enough for real-time applications; and 3) the computation of mvMSE for signals with a large number of channels requires the storage of a huge number of elements. \textit{Methods}: To deal with these problems and improve the stability of mvMSE, we introduce multivariate multiscale dispersion entropy (MDE - mvMDE), as an extension of our recently developed MDE, to quantify the complexity of multivariate time series. \textit{Results}: We assess mvMDE, in comparison with mvMSE and multivariate multiscale fuzzy entropy (mvMFE), on correlated and uncorrelated multi-channel noise signals, bivariate autoregressive processes, and three biomedical datasets. The results show that mvMDE takes into account dependencies in patterns across both the time and spatial domains. The mvMDE, mvMSE, and mvMFE methods are consistent in that they lead to similar conclusions about the underlying physiological conditions. However, the proposed mvMDE discriminates various physiological states of the biomedical recordings better than mvMSE and mvMFE. In addition, for both the short and long time series, the mvMDE-based results are noticeably more stable than the mvMSE- and mvMFE-based ones. \textit{Conclusion}: For short multivariate time series, mvMDE, unlike mvMSE, does not result in undefined values. Furthermore, mvMDE is noticeably faster than mvMFE and mvMSE and also needs to store a considerably smaller number of elements. \textit{Significance}: Due to its ability to detect different kinds of dynamics of multivariate signals, mvMDE has great potential to analyse various physiological signals.
\end{abstract}


\IEEEpeerreviewmaketitle

\section{Introduction}
\IEEEPARstart{M}{ultivariate} techniques are needed to analyse data consisting of more than one time series \cite{cerutti2009multiscale}. The majority of physiological and pathophysiological activities include interactions between different kinds of single processes. Thus, we expect that parameters or measures with different origins are considered in a multivariate way \cite{cerutti2009multiscale,cerutti2012multivariate}. Furthermore, recent developments in sensor technology enabling routine recordings of multi-channel signals have led to an increasing popularity of this kind of analysis on biomedical data \cite{ahmed2012multivariate,cerutti2009multiscale}.

Advances on information theory and non-linear dynamical approaches have recently allowed the study of different kinds of multivariate time series \cite{pereda2005nonlinear}. Due to the intrinsic non-linearity of diverse physiological processes, non-linear analysis of multivariate time series has been broadly used in biomedical engineering with the aim of studying the relationship between simultaneously recorded signals \cite{pereda2005nonlinear}.

Multivariate multiscale entropy (mvMSE) as a powerful non-linear measure is based on a combination of multivariate sample entropy (SampEn - mvSE) and the coarse-graining process \cite{ahmed2011multivariate}. mvMSE, by taking into account both the spatial and time domains, shows the complexity of multi-channel signals \cite{ahmed2011multivariate}. Complexity reflects the degree of structural richness of time series \cite{costa2005multiscale,ahmed2011multivariate} and is different with that of irregularity or uncertainty defined from classical entropy methods such as SampEn \cite{richman2000physiological}, permutation entropy (PerEn) \cite{bandt2002permutation}, and dispersion entropy (DisEn) \cite{rostaghi2016dispersion}. That is to say, neither completely regular or certain nor completely irregular (uncorrelated random) time series are truly complex, since none of them is structurally rich at a global level \cite{fogedby1992phase,costa2005multiscale,silva2015multiscale,ahmed2011multivariate}.

The multivariate multiscale entropy-based analysis is interpreted based on: 1) the multivariate time series \textbf{X} is more complex than the multivariate time series \textbf{Y}, if for the most temporal scales, the mvSE measures for \textbf{X} are larger than those for \textbf{Y}; 2) a monotonic fall in the multivariate entropy values along the temporal scale factors shows that the signal only includes useful information at the smallest scale factors; and 3) a multivariate signal illustrating long-range correlations and complex creating dynamics is characterized by either a constant mvSE or this demonstrates a monotonic rise in mvSE with the temporal scale factor \cite{ahmed2011multivariate}.

Although the mvMSE is a powerful and widely-used method, when applied to short signals, the results may be undefined or unreliable \cite{azami2017refined}. To alleviate this shortcoming, multivariate multiscale fuzzy entropy (mvMFE) based on multivariate fuzzy entropy (mvFE) and the coarse-graining process was suggested \cite{li2014coupling}. To decrease the running time of the mvMFE proposed in \cite{li2014coupling}, we have recently proposed an mvMFE with a new fuzzy membership function \cite{azami2017refined}. Nevertheless, the mvMFE is still slow for real-time applications and may lead to unreliable results for short signals, as shown later.

To overcome the problem of unreliable values for mvMFE and mvMSE, multivariate multiscale PerEn (mvMPE) was proposed \cite{morabito2012multivariate}. To have more information regarding the amplitude of multi-channel signals, multivariate weighted multiscale PerEn (mvWMPE) has recently been developed \cite{yin2017multivariate}. However, both the mvMPE and mvWMPE do not take into account the cross-statistical properties between multiple input channels and do not follow the concept of complexity for some signals such as white Gaussian noise (WGN) and $1/f$ noise \cite{azami2017refined,fogedby1992phase,ahmed2011multivariate}.

mvMSE and mvMFE have growing appeal and broad use. They have been successfully used in a number of biomedical signal processing applications, such as, to characterise electroencephalogram (EEG) signals in Alzheimer's disease (AD) \cite{azami2017univariate,labate2013entropic}, to analyze the multivariate cardiovascular time series \cite{zhao2016multivariable}, to characterize focal and non-focal EEG time series \cite{azami2017refined}, to analyze the complexity of interbeat interval and interbreath signals \cite{ahmed2011multivariate}, and to analyze the postural fluctuations in fallers and non-fallers older adults \cite{ramdani2016parameters}. 

However, mvMSE and mvMFE have the following shortcomings: 1) mvMSE and mvMFE values may be unreliable and unstable for short signals; 2) they are not quick enough for real-time applications; and 3) computation of mvMSE and mvMFE of a signal with a large number of channels needs to have large memory space, as shown later. To address these drawbacks, we propose four algorithms to extend our recently developed MDE to its multivariate versions, termed multivariate MDE (mvMDE). To evaluate the mvMDE methods, we use both synthetic and real multivariate datasets. Our results indicate that mvMDE is noticeably faster than the existing methods, leads to more stable results, better discriminates different kinds of biomedical time series, does not lead to undefined values for short multivariate time series, and needs to store a considerably smaller number of elements in comparison with mvMSE and mvMFE.


\section{Multivariate Multiscale Dispersion Entropy (mvMDE)}
In this study, we propose and explore three different alternative implementations of mvMDE until we arrive at a  fourth and preferred one. All the mvMDE implementations include two main steps: 1) coarse-graining process for multivariate time series; and 2) multivariate dispersion entropy (mvDE), as an extension of our recently developed DisEn \cite{rostaghi2016dispersion}. It is worth noting that for all the mvMDE algorithms, the mapping based on the normal cumulative distribution function (NCDF) used in the calculation of mvDE for the first temporal scale factor is maintained fixed across all scales. In fact, in the mvMDE, $\mu$ and $\sigma$ of the NCDF are respectively set at the average and standard deviation (SD) of the original time series and they remain constant for all temporal scale factors. This fact is similar to $r$ in the mvMSE and mvMFE, setting at a certain percentage (usually 15\%) of the SD of the original signal and remaining constant for all scales \cite{ahmed2011multivariate,azami2017refined}. 

\subsection{Coarse-Graining Process for Multivariate Signals}
Assume we have a \textit{p}-channel time series $ \mathbf{U}=\{u_{k,b}\}_{k=1,2,\dotso,p}^{b=1,2,\dotso,L}$ of length $L$. In the mvMDE algorithms, for each channel, the original signal is first divided into non-overlapping segments of length $\tau$, named scale factor. Next, for each channel, the average of each segment is calculated to derive the coarse-grained signals as follows \cite{ahmed2011multivariate,azami2017refined}:
\begin{equation} 
{{x}_{k,i}}^{(\tau)}=\frac{1}{\tau}\sum\limits_{b=(i-1)\tau +1}^{i\tau}{{{u}_{k,b}}}, \: 1\le i\le \left\lfloor \frac{L}{\tau } \right\rfloor =N\ ,\: 1\le k\le p
\end{equation}
where $ N $ denotes the length of the coarse-grained signal. The second step of mvMDE is calculating the mvDE of each coarse-grained signal.
\subsection {Background Information for the mvDE}
We build four diverse alternative implementations of mvDE (mvDE-I to III and mvDE) until we arrive at a preferred (or optimal) one, i.e., mvDE. However, we here present all the simpler alternatives (mvDE-I to III), since they can still be useful in some settings and allow for clearer comparisons with other current approaches.

\subsubsection{mvDE-I}
The mvDE-I of the multi-channel coarse-grained time series $ \mathbf{X}=\{x_{k,i}\}_{k=1,2,\dotso,p}^{i=1,2,\dotso,N}$, which is based on the mvMPE algorithm \cite{morabito2012multivariate}, is calculated as follows:

\textit{a}) First, $\mathbf{X}=\{x_{k,i}\}_{k=1,2,\dotso,p}^{i=1,2,\dotso,N}$ are mapped to $\textit{c}$ classes with integer indices from 1 to $\textit{c}$. Since the amplitude values of each of series $\mathbf{x}_{k}$ ($k=1,2,\dotso,p$) may be dominated by the components of vectors coming from the time series with the largest amplitudes, we scale all of the data channels to the same amplitude range. To this end and to overcome the problem of assigning the majority of $x_{k,i}$ to only few classes when maximum or minimum values are noticeable larger or smaller than the mean/median value of the signal, the NCDF of each of $\textbf{x}_k$ is first calculated. In fact, the NCDF maps $\textbf{X}$ into ${\textbf{Y}}=\{y_{k,i}\}_{k=1,2,\dotso,p}^{i=1,2,\dotso,N}$ from 0 to 1 as follows: 

\begin{equation}
y_{k,i}=\frac{1}{\sigma_k\sqrt{2\pi}}\int\limits_{-\infty}^{x_{k,i}}{e}^{\frac{-(t-\mu_k)^2}{2\sigma_k^2}}\,\mathrm{d}t
\end{equation}

\setlength{\parindent}{0pt}%
where $\sigma_k$ and $\mu_k$ are the SD and mean of time series $\textbf{x}_k$, respectively. Then, we use a linear algorithm to assign each $y_{k,i}$ to an integer from 1 to $c$. To do so, for each member of the mapped signal, we use $z_{k,i}^{c}=\mbox{round}(c\cdot y_{k,i}+0.5)$, where  $z_{k,i}^{c}$ denotes the $i^{th}$ member of the classified signal in the $k^{th}$ channel and rounding involves either increasing or decreasing a number to the next digit. Note that, although this part is linear, the whole mapping approach is non-linear because of the use of NCDF.
\setlength{\parindent}{9pt}%

\textit{b}) Time series $\textbf{z}^{m,c}_{k,j}$ are made with embedding dimension $m$ and time delay $d$ according to $\textbf{z}^{m,c}_{k,j}=\{z^{c}_{k,j},{z}^{c}_{k,j+d},+\dots+{z}^{c}_{k,j+(m-1)d}\} $, $j=1,2,\dots,N-(m-1)d$ \cite{rostaghi2016dispersion}\cite{richman2000physiological}\cite{bandt2002permutation}. Each time series $\textbf{z}^{m,c}_{k,j}$ is mapped to a dispersion pattern ${{\pi }_{{{v}_{0}}{{v}_{1}}\dots{{v}_{m-1}}}}$, where $ z_{k,j}^{c}={{v}_{0}} $ ,$ z_{k,j+d}^{c}={{v}_{1}} $ ,\dots, $z_{k,j+(m-1)d}^{c}={{v}_{m-1}}$. The number of possible dispersion patterns that can be assigned to each time series $\textbf{z}^{m,c}_{k,j}$ is equal to $c^{m}$, since the signal has $m$ members and each member can be one of the integers from 1 to $c$ \cite{rostaghi2016dispersion}. 

\textit{c}) For each channel $1\leq k\leq p$ and for each of $c^{m}$ potential dispersion patterns ${{\pi }_{{{v}_{0}}\dots{{v}_{m-1}}}}$, relative frequency is obtained as follows:
\begin{equation}
\begin{split}
p({{\pi }_{{{v}_{0}}\dots{{v}_{m-1}}}})=\\
\frac{\#\{j\left| j\le N-(m-1)d,\mathbf{z}_{k,j}^{m,c}\text{ has type }{{\pi }_{{{v}_{0}}\dots{{v}_{m-1}}}} \right.\}}{(N-(m-1)d)p}
\end{split}
\end{equation}
\setlength{\parindent}{0pt}%
where $\#$ means cardinality. In fact, $p({{\pi }_{{{v}_{0}}\dots{{v}_{m-1}}}})$ shows the number of dispersion patterns of ${{\pi }_{{{v}_{0}}\dotso{{v}_{m-1}}}}$ that is assigned to $\textbf{z}^{m,c}_{k,j}$, divided by the total number of embedded signals with embedding dimension $m$ multiplied by the number of channels.
\setlength{\parindent}{9pt}%

\textit{d}) Finally, based on the Shannon's definition of entropy, the mvDE-I is calculated as follows:
\begin{equation}
\begin{split}
mvDE\!-\!I(\mathbf{X},m,c,d)=\\
-\sum_{\pi=1}^{c^m} {p({{\pi}_{{{v}_{0}}\dots{{v}_{m-1}}}})\cdot\ln }\left(p({{\pi}_{{{v}_{0}}\dots{{v}_{m-1}}}}) \right)      
\end{split}
\end{equation}

In case all possible dispersion patterns have equal probability value, the highest value of mvDE-I is obtained, which has a value of $ln ({{c}^{m}})$. In contrast, if there is only one $ p({{\pi }_{{{v}_{0}}\dots{{v}_{m-1}}}}) $ different from zero, which demonstrates a completely regular/certain signal, the smallest value of mvDE-I is obtained. In the algorithm of mvDE-I, we compare $Np$ dispersion patterns of a \textit{p}-channel signal with $c^m$ potential patterns. Thus, at least $c^m+Np$ elements are stored. 

To work with reliable statistics to calculate MDE, it was recommended $c^{m}<\left\lfloor \frac{L}{\tau_{max}} \right\rfloor$ \cite{azami2017refineddd}. For mvMDE-I, since the mvDE-I counts the dispersion patterns for every channel of a multivariate time series, it is suggested $c^{m}<\left\lfloor \frac{pL}{\tau_{max} } \right\rfloor$. mvMDE-I works appropriately when the components of a multivariate signal are statistically independent. However, the mvMDE-I algorithm, like mvMPE \cite{morabito2012multivariate}, does not consider the spatial domain of time series. To overcome this problem, we propose mvMDE-II based on the Taken's theorem \cite{cao1998dynamics,azami2017refined}.

\subsubsection {mvDE-II}

The algorithm of mvDE-II is as follows:

\textit{a}) First, like mvDE-I, $\mathbf{X}=\{x_{k,i}\}_{k=1,2,\dotso,p}^{i=1,2,\dotso,N}$ are mapped to $\textbf{Z}=\{z_{k,i}\}_{k=1,2,\dotso,p}^{i=1,2,\dotso,N}$ based on the NCDF.

\textit{b}) To take into account both the spatial and time domains, multi-channel embedded vectors are generated according to the multivariate embedding theory \cite{cao1998dynamics}. The multivariate embedded reconstruction of \textbf{Z} is defined as:
\begin{equation}
\begin{split}
Z_\textbf{m}(j)=[z_{1,j},z_{1,j+d_1},\dotso,z_{1,j+(m_1-1)d_1},\\
z_{2,j},z_{2,j+d_2},\dotso,z_{2,j+(m_2-1)d_2},\dotso,\\
z_{p,j},z_{p,j+d_p},\dotso,z_{p,j+(m_p-1)d_p}]
\end{split}
\end{equation}
 \setlength{\parindent}{0pt}%
where $\textbf{m}=[m_1,m_2,\dotso,m_p]$ and $\textbf{d}=[d_1,d_2,\dotso,d_p]$ denote the embedding dimension and the time lag vectors, respectively. Note that the length of $Z_\textbf{m}(j)$ is $\sum_{k=1}^{p} m_k$. For simplicity, we assume $d_k=d$ and $m_k=m$, that is, all the embedding dimension values and all the delay values are equal. 
\setlength{\parindent}{9pt}%

\textit{c}) Each series $Z_\textbf{m}(j)$ is mapped to a dispersion pattern ${{\pi}_{{{v}_{0}}{{v}_{1}}\dots{{v}_{mp-1}}}}$, where $ z_{1,j}^{c}={{v}_{0}}$, $ z_{1,j+d}^{c}={{v}_{1}}$,\dots, $z_{p,j+(m-1)d}={{v}_{mp-1}}$. The number of possible dispersion patterns that can be assigned to each time series $Z_\textbf{m}(j)$ is equal to $c^{mp}$, since the signal has $mp$ members and each member can be one of the integers from 1 to $c$.

\textit{d}) For each of $c^{mp}$ potential dispersion patterns ${{\pi }_{{{v}_{0}}\dots{{v}_{mp-1}}}}$, relative frequency is obtained based on the DisEn algorithm \cite{rostaghi2016dispersion} as follows:
\begin{equation}
\begin{split}
p({{\pi }_{{{v}_{0}}\dots{{v}_{mp-1}}}})=\\
\frac{\#\{j\left| j\le N-(m-1)d,Z_\textbf{m}(j)\text{ has type }{{\pi }_{{{v}_{0}}\dots{{v}_{mp-1}}}} \right.\}}{N-(m-1)d}
\end{split}
\end{equation}

\textit{e}) Finally, based on the Shannon's definition of entropy, the mvDE-II is calculated as follows:
\begin{equation}
\begin{split}
mvDE\!-\!II(\mathbf{X},m,c,d)=\\
-\sum_{\pi=1}^{c^{mp}} {p({{\pi}_{{{v}_{0}}\dots{{v}_{mp-1}}}})\cdot\ln }\left(p({{\pi}_{{{v}_{0}}\dots{{v}_{mp-1}}}}) \right)      
\end{split}
\end{equation}

In the algorithm of mvDE-II, at least $c^{mp}+Np$ elements are stored. Thus, when \textit{p} is large, the algorithm needs huge space of memory to store elements. To work with reliable statistics to calculate mvMDE-II, it is recommended $c^{mp}<\left\lfloor \frac{L}{\tau_{max} } \right\rfloor$. Thus, although mvDE-II deals with both the spatial and time domains, the length of a signal and its number of channels should be very large and small, respectively, to reliably calculate mvDE-II values. To alleviate the problem, we propose mvDE-III.

\subsubsection {mvDE-III}

The algorithm of mvDE-III is as follows:

\textit{a}) First, like the mvDE-I and mvDE-II approaches, $\mathbf{X}=\{x_{k,i}\}_{k=1,2,\dotso,p}^{i=1,2,\dotso,N}$ are mapped to $\textbf{Z}=\{z_{k,i}\}_{k=1,2,\dotso,p}^{i=1,2,\dotso,N}$.

\textit{b}) Multivariate embedded vectors $\textbf{Z}_{k,\textbf{m}}(j)$ are generated according to the Taken's embedding theorem \cite{cao1998dynamics} with \textit{p} embedding dimension vectors $\textbf{m}_k=[1,1,\dotso,m_k,\dotso,1,1]$ ($k=1,\dots,p$) with length $m+p-1$, where $m_k$ denotes the $k^{th}$ element of \textbf{m}. For simplicity, we assume $m_k=m$ and $d_k=d$. 

\textit{c}) Each series $\textbf{Z}_{k,\textbf{m}}(j)$ is mapped to a dispersion pattern ${{\pi}_{{{v}_{0}}{{v}_{1}}\dots{{v}_{m+p-2}}}}$. The number of possible dispersion patterns that can be assigned to each time series $\textbf{Z}_{k,\textbf{m}}(j)$ is equal to $c^{m+p-1}$, since the signal has $m+p-1$ members and each member can be one of the integers from 1 to $c$ \cite{rostaghi2016dispersion}. Since we count the number of patterns for each of \textit{p} different $\textbf{m}_k$, we have a $p.c^{mp-m-p+1}$ times increase in the number of dispersion patterns in comparison with mvDE-II, leading to more reliable results for a signal with a small number of sample points, as shown later.

\textit{d}) For each channel $1\leq k\leq p$ and for each of $c^{m+p-1}$ potential dispersion patterns ${{\pi }_{{{v}_{0}}\dots{{v}_{m+p-2}}}}$, relative frequency is obtained as follows:
\begin{equation}
\begin{split}
p({{\pi }_{{{v}_{0}}\dots{{v}_{m+p-2}}}})=\\
\frac{\#\{j\left| j\le N-(m-1)d,\textbf{Z}_{k,\textbf{m}}(j)\text{ has type }{{\pi }_{{{v}_{0}}\dots{{v}_{m+p-2}}}} \right.\}}{(N-(m-1)d)p}
\end{split}
\end{equation}

\textit{e}) Finally, based on the Shannon's definition of entropy, the mvDE-II is calculated as follows:
\begin{equation}
\begin{split}
mvDE\!-\!III(\mathbf{X},m,c,d)=\\
-\sum_{\pi=1}^{c^{m+p-1}} {p({{\pi}_{{{v}_{0}}\dots{{v}_{m+p-2}}}})\cdot\ln }\left(p({{\pi}_{{{v}_{0}}\dots{{v}_{m+p-2}}}}) \right)      
\end{split}
\end{equation}

In the algorithm of mvDE-III, at least $c^{m+p-1}+Np$ elements are stored. Although this number is noticeably smaller than that for mvDE-II, the algorithm still needs to have large memory space for a signal with a large number of channels. To work with reliable statistics to calculate mvMDE-III, it is recommended $c^{m+p-1}<\left\lfloor \frac{pL}{\tau_{max} } \right\rfloor$. Therefore, albeit mvMDE-III takes into account both the spatial and time domains and needs to smaller number of sample points in comparison with mvMDE-II, there is a need to have a large enough number of samples and small number of channels. To alleviate these deficiencies, we propose mvDE.

\subsection {Multivariate Dispersion Entropy (mvDE)}
The mvDE algorithm is as follows:

\textit{a}) First, like mvDE-I to III, the multivariate signal $\mathbf{X}=\{x_{k,i}\}_{k=1,2,\dotso,p}^{i=1,2,\dotso,N}$ is mapped to $\textit{c}$ classes with integer indices from 1 to $\textit{c}$.

\textit{b}) Like mvDE-II, to consider both the spatial and time domains, multivariate embedded vectors $Z_\textbf{m}(j), 1\leq j\leq N-(m-1)d$ are created based on the Taken's embedding theorem \cite{cao1998dynamics}. For simplicity, we assume $d_k=d$ and $m_k=m$. 

\textit{c}) For every $Z_\textbf{m}(j)$, all combinations of the $\sum_{k=1}^{p}m_k$ elements in $Z_\textbf{m}(j)$ taken \textit{m} at a time, termed $\phi_q(j)$ ($q=1,...\binom{mp}{m}$), are created. The number of the combinations is equal to $\binom{mp}{m}$. Therefore, for all channels, we have $(N-(m-1)d)\binom{mp}{m}$ dispersion patterns.

\textit{d}) For each $1\leq q\leq\binom{mp}{m}$ and for each of $c^{m}$ potential dispersion patterns ${{\pi }_{{{v}_{0}}\dots{{v}_{m-1}}}}$, relative frequency is obtained as follows:
\begin{equation}
\begin{split}
p({{\pi }_{{{v}_{0}}\dots{{v}_{m-1}}}})=\\
\frac{\#\{j\left| j\le N-(m-1)d,\phi_q(j)\text{ has type }{{\pi }_{{{v}_{0}}\dots{{v}_{m-1}}}} \right.\}}{(N-(m-1)d)\binom{mp}{m}}
\end{split}
\end{equation}

\textit{e}) Finally, based on the Shannon's definition of entropy, the mvDE is calculated as follows:
\begin{equation}
\begin{split}
mvDE(\mathbf{X},m,c,d)=\\
-\sum_{\pi=1}^{c^m} {p({{\pi}_{{{v}_{0}}\dots{{v}_{m-1}}}})\cdot\ln }\left(p({{\pi}_{{{v}_{0}}\dots{{v}_{m-1}}}}) \right)      
\end{split}
\end{equation}  

In the mvDE algorithm, at least $c^{m}+Np$ elements are stored. This number is noticeably smaller than those for mvDE-I to III, leading to more stable results for signals with a short length and a large number of samples. As the number of patterns obtained by the mvMDE method is $(N-(m-1)d)\binom{mp}{m}$, it is suggested $c^{m}<\left\lfloor \frac{L\binom{mp}{m}}{\tau_{max} } \right\rfloor$ to work with reliable statistics.

\subsection{Parameters of the mvMDE, mvMSE, and mvMFE Methods}
In addition to the maximum scale factor $\tau_{max}$ described before, there are three other parameters for the mvMDE methods, including the embedding dimension vector \textbf{m}, the number of classes \textit{c}, and the time delay vector $\textbf{d}$. In practice, it is recommended $ d_k=1 $, because some information with regard to the frequency may be ignored for $ 1<d $. We need $1<c$ to keep away the trivial case of having only one dispersion pattern. For simplicity, we use $c=5$ and $m_k=2$ for all signals used in this study, although the range $2<c<9$ leads to similar results. For more information about \textit{c}, $m_k$, and $d_k$, please refer to \cite{rostaghi2016dispersion}.

In this study, $d_k$, $ m_k $, and $r$ for the mvMSE and mvMFE were respectively set as 1, 2, and 0.15 of the SD of the original time series following recommendations in \cite{ahmed2011multivariate,azami2017refined}. The maximum scale factor for mvMSE and mvMFE also follows \cite{azami2017refined,ahmed2011multivariate}. In the algorithm of mvSE and mvFE, at least $\binom{Np}{2}+Np(pm+1)$ elements are stored. Matlab codes of mvMFE and mvMSE are available at http://dx.doi.org/10.7488/ds/1432. Overall, the characteristics and limitations of the mvSE, mvFE, and mvDE algorithms for a \textit{p}-channel signal with length \textit{N} are summarized in Table I.

\begin{table*}
		\setlength{\tabcolsep}{3pt}
		\renewcommand{\arraystretch}{1.4}
	\centering
	\label{tab:table4}\caption{Ability to deal with spatial domain and characterization of short signals, minimum number of elements to be stored, and minimum number of samples needed for each of the mvSE, mvFE, and mvDE algorithms for a \textit{p}-channel signal with length \textit{N}.} 
	\begin{tabular}{c*{8}{c}}
		Methods   &Spatial domain& Short signals& Minimum number of elements to be stored& Minimum number of samples \\
		\hline
		mvSE \cite{ahmed2012multivariate} & yes& unreliable or undefined& $\binom{Np}{2}+Np(pm+1)$  & $10^{m}<N$   \\
		mvFE  \cite{azami2017refined}& yes & unreliable&$\binom{Np}{2}+Np(pm+1)$ & $10^{m}<N$ \\
		mvPE \cite{morabito2012multivariate} and mvWPE \cite{yin2017multivariate}& no &reliable&$m!+Npm$& $m!<N$ \\
		mvDE-I  & no & reliable& $c^m+Np$& $\frac{c^{m}}{p}<N$\\
		mvDE-II  & yes & unreliable& $c^{mp}+Np$ &$c^{mp}<N$\\
		mvDE-III & yes &unreliable& $c^{m+p-1}+Np$&$\frac{c^{m+p-1}}{p}<N$\\			
		mvDE  & yes & reliable&  $c^{m}+Np$ &$\frac{c^{m}}{\binom{mp}{m}}<N$\\
		
	\end{tabular}
\end{table*}

\section{Evaluation Signals}
In this section, the descriptions of correlated and uncorrelated noise signals and real time series used in this study are given.

\subsection{Synthetic Signals}
The irregularity of multivariate $ 1/f $ noise is lower than multivariate WGN, whereas the complexity of the former is higher than the latter \cite{fogedby1992phase,azami2017refined,ahmed2011multivariate}. Thus, $1/f$ noise and WGN signals have been commonly used to assess the multivariate multiscale entropy techniques \cite{humeau2016multivariate,azami2017refined,ahmed2011multivariate}. For more information, please refer to \cite{costa2005multiscale,fogedby1992phase,azami2017refined,ahmed2011multivariate}.

To understand the behaviour of the mvMDE methods on uncorrelated WGN and $1/f$ noise, we first generated a trivariate time series, where originally all three data channels were realization of mutually independent $1/f$ noise. Then, we gradually decreased the number of data channels representing $1/f$ noise (from 3 to 0) and at the same time, increased the number of variates representing independent WGN (from 0 to 3) \cite{humeau2016multivariate}. The number of channels was always three.

To create correlated bivariate noise time series, we first generated a bivariate uncorrelated random time series \textbf{H}. Afterwards, \textbf{H} was multiplied with the standard deviation (hereafter, sigma) and then, the value of the mean (hereafter, mu) was added. Next, \textbf{H} was multiplied by the upper triangular matrix \textbf{L} obtained from the Cholesky decomposition of a defined correlation matrix \textbf{R} (which is positive and symmetric) to set the correlation. Here, we set $\textbf{R}=\begin{bmatrix}
1 & 0.95 \\[0.3em]
0.95 & 1
\end{bmatrix}$ according to \cite{ahmed2011multivariate,azami2017refined}. An in-depth study on the effect of correlated and uncorrelated $1/f$ noise and WGN on multiscale entropy approaches can be found in \cite{costa2005multiscale,ahmed2011multivariate}. 

Based on the fact that the larger the order of an AR process, the more complex the AR process \cite{ahmed2011multivariate}, we evaluate the mvMDE, mvMSE, and mvMFE methods on a bivariate AR process describing the evolution of a set of two variables as a linear function of their past values according to:

\begin{equation}
\textbf{y}_n=\textbf{e}_n+\sum_{\gamma=1}^{\alpha}{\textbf{y}_{n-\gamma}\textbf{A}_\gamma}      
\end{equation}

\setlength{\parindent}{0pt}%
where \textbf{y} is the $2\times 1$ vector of variables, $\alpha$ is the maximum lag in the bivariate AR model, $\mathbf{A}_\gamma$ denotes the $2\times 2$ matrix of parameters corresponding to lag order $\gamma$, and $\textbf{e}_n$ is the $2\times 1$ vector of error terms assumed to be WGN\cite{penny2002bayesian}. For simplicity, we set $A_{\gamma}=\begin{bmatrix}
0.15 & 0.15 \\[0.3em]
0.15 & 0.15
\end{bmatrix}$.
\setlength{\parindent}{9pt}%

\subsection{Real Biomedical Datasets}
\textit{1) Dataset of Stride Internal Fluctuations}: To investigate the ability of the proposed mvMDE methods to reveal the long-range correlations and dynamics of multivariate signals, the stride interval recordings are used \cite{ahmed2012dynamical,hausdorff1996fractal}. The time series were recorded from ten young, healthy men. Mean age was 21.7 years, changing from 18 to 29 years. Height and weight were 1.77 $\pm$ 0.08 meters (mean $\pm$ SD) and 71.8 $\pm$ 10.7 kg (mean $\pm$ SD), respectively. All ten participants provided informed written consent walking for 1 hour at slow, normal, and fast paces and also walking a metronome set to each subject's mean stride interval. Three walking paces were considered as different variables from the same system. In this way, we expect to be able to discriminate between the metronomically-paced and self-spaced walking. For further information, please refer to \cite{hausdorff1996fractal}.

\textit{2) Dataset of Focal and Non-focal Brain Activity}: The ability of the mvMDE methods, in comparison with mvMFE and mvMSE, to differentiate focal from non-focal recordings is evaluated using a publicly-available EEG dataset \cite{andrzejak2012nonrandomness}. The dataset includes 5 patients and, for each patient, there are 750 focal and 750 non-focal bivariate signals. The length of each recording was 20 s with sampling frequency of 512 Hz (10240 sample points). Further information can be found in \cite{andrzejak2012nonrandomness}. Before computing the aforementioned methods, all recordings were digitally filtered employing an FIR band-pass filter with cut-off frequencies at 0.5 Hz and 40 Hz.

\textit{3) Surface MEG Recordings in Alzheimer's Disease}: We analysed resting state MEG time series recorded with a 148-channel whole-head magnetometer. All 62 participants agreed for the research, which was approved by the local ethics committee. To screen the cognitive status, a mini-mental state examination (MMSE) was done. There were 36 AD patients (age = $74.06 \pm 6.95$ years, all data given as mean $\pm$ SD, and MMSE score = $18.06 \pm 3.36$) and 26 controls (age = $71.77 \pm 6.38$ years, and MMSE score = $28.88 \pm 1.18$). The difference in age between two groups was not significant ($p$-value = 0.1911, Student's $t$-test) \cite{escudero2011regional}. The distribution of MEG sensors is shown in Fig. 2 in \cite{escudero2011regional}. For each participant, five minutes of MEG resting state activity were recorded at a sampling frequency of 169.5 Hz. The signals were divided into 10 s segments (1695 samples) and visually inspected using an automated thresholding procedure to discard epochs noticeably contaminated with artifacts. All recordings were digitally band-pass filtered with a Hamming window FIR filter of order 200 and cut-off frequencies at 1.5 Hz and 40 Hz. For more information, please see \cite{escudero2011regional}.

\section{Results and Discussions}
\subsection{Synthetic Signals}
We first apply the proposed and existing methods to 40 independent realizations of uncorrelated trivariate WGN and $1/f$ noise, described in Section III. The number of sample points for each of the $ 1/f $ noise and WGN signals were 15000. The average and SD of the results for mvMDE-I, mvMDE-II, mvMDE-III, mvMDE, mvMSE, and mvMFE are depicted in Fig. 1(a) to 1(f), respectively. Using all the existing and proposed methods, the entropy values of trivariate WGN signals are higher than those of the other trivariate time series at low scale factors. However, the entropy values for the coarse-grained trivariate $1/f$ noise signals stay almost constant or decrease slowly along the temporal scale factor, while the entropy values for the coarse-grained WGN signal monotonically decreases with the increase of scale factors. When the length of WGN signals, obtained by the coarse-graining process, decreases (i.e., the scale factor increases), the mean value of inside each signal converges to a constant value and the SD becomes smaller. Therefore, no new structures are revealed at higher temporal scales. This demonstrates a multivariate WGN time series has information only in small temporal scale factors. In contrast, for trivariate $ 1/f $ noise signals, the mean value of the fluctuations inside each signal does not converge to a constant value.

For all the methods, the higher the number of variates representing $1/f$ noise, the higher complexity the trivariate signal, in agreement with the fact that multivariate $1/f$ noise is structurally more complex than multivariate WGN \cite{ahmed2011multivariate,azami2017refined,fogedby1992phase}. Here, for multivariate $1/f$ noise and WGN, $\tau_{max}$ was 20 for mvMDE, according to Section II.

\begin{figure*}
	\centering
	\begin{multicols}{3}
		\includegraphics[width=6.2cm,height=3.5cm]{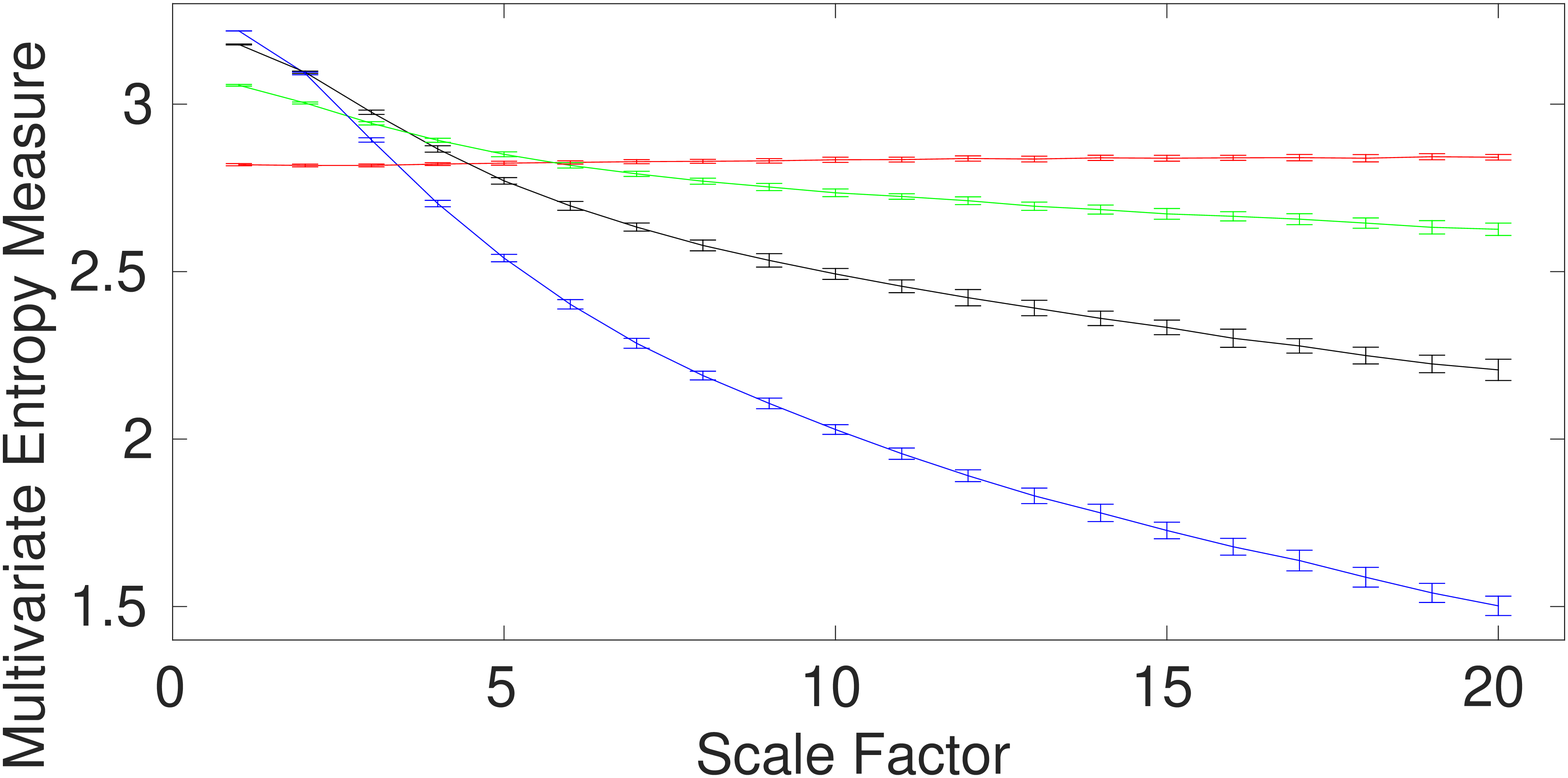}
		\footnotesize{(a) mvMDE-I}
		\includegraphics[width=6.2cm,height=3.5cm]{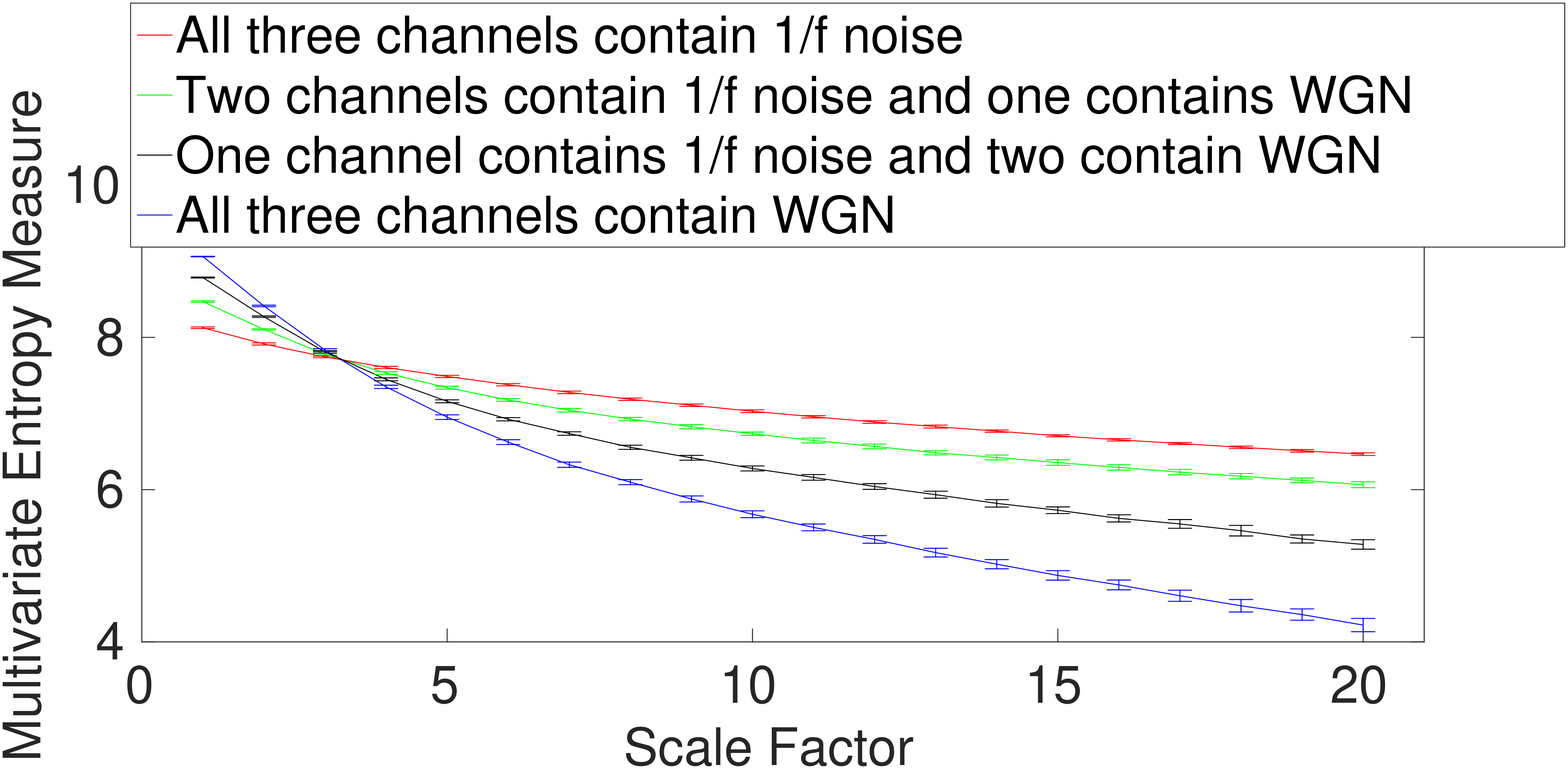}
		\footnotesize{(b) mvMDE-II}
		\includegraphics[width=6.2cm,height=3.5cm]{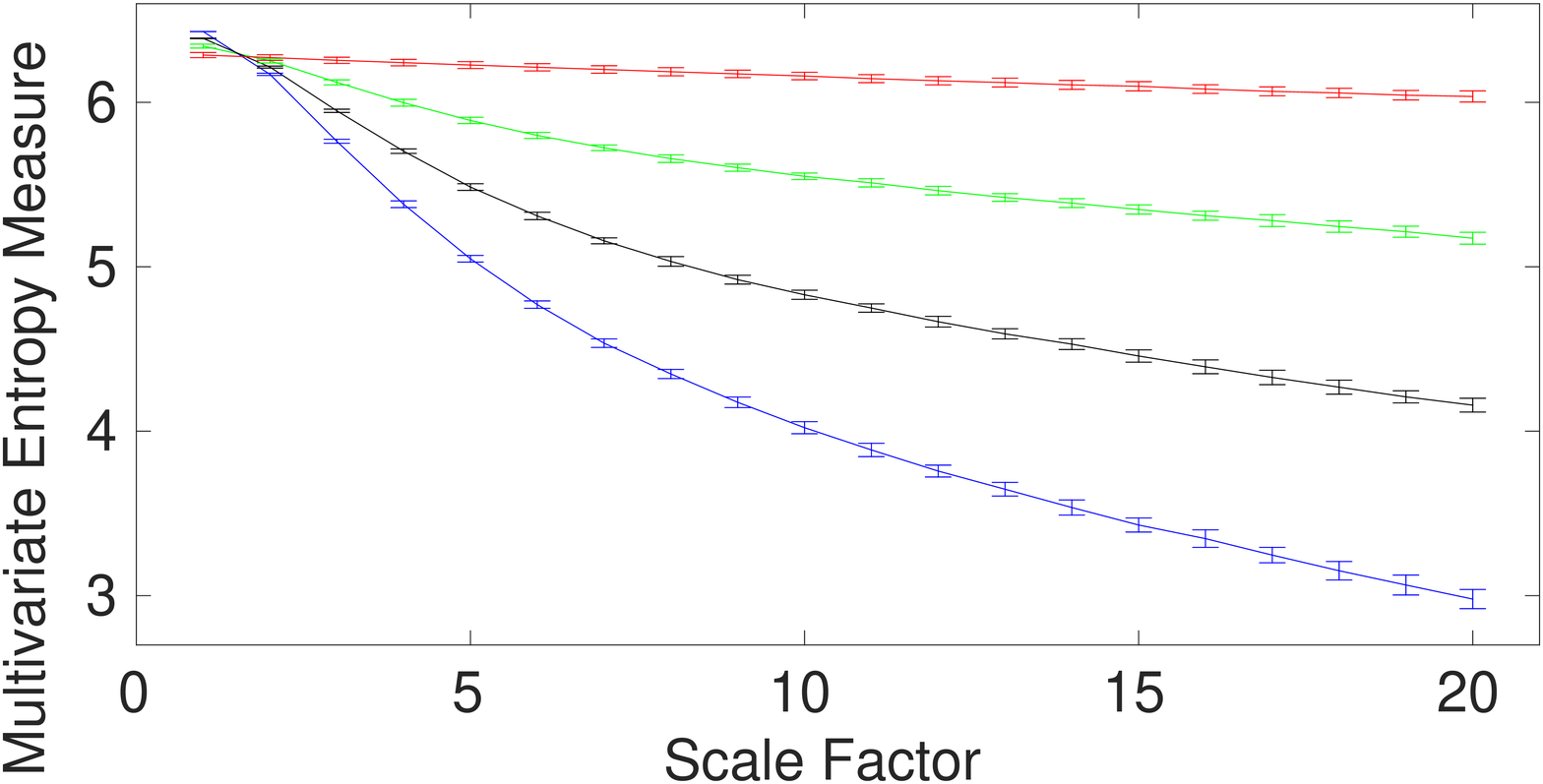}
		\footnotesize{(c) mvMDE-III}
	\end{multicols}
	\begin{multicols}{3}
		\centering
		\includegraphics[width=6.2cm,height=3.5cm]{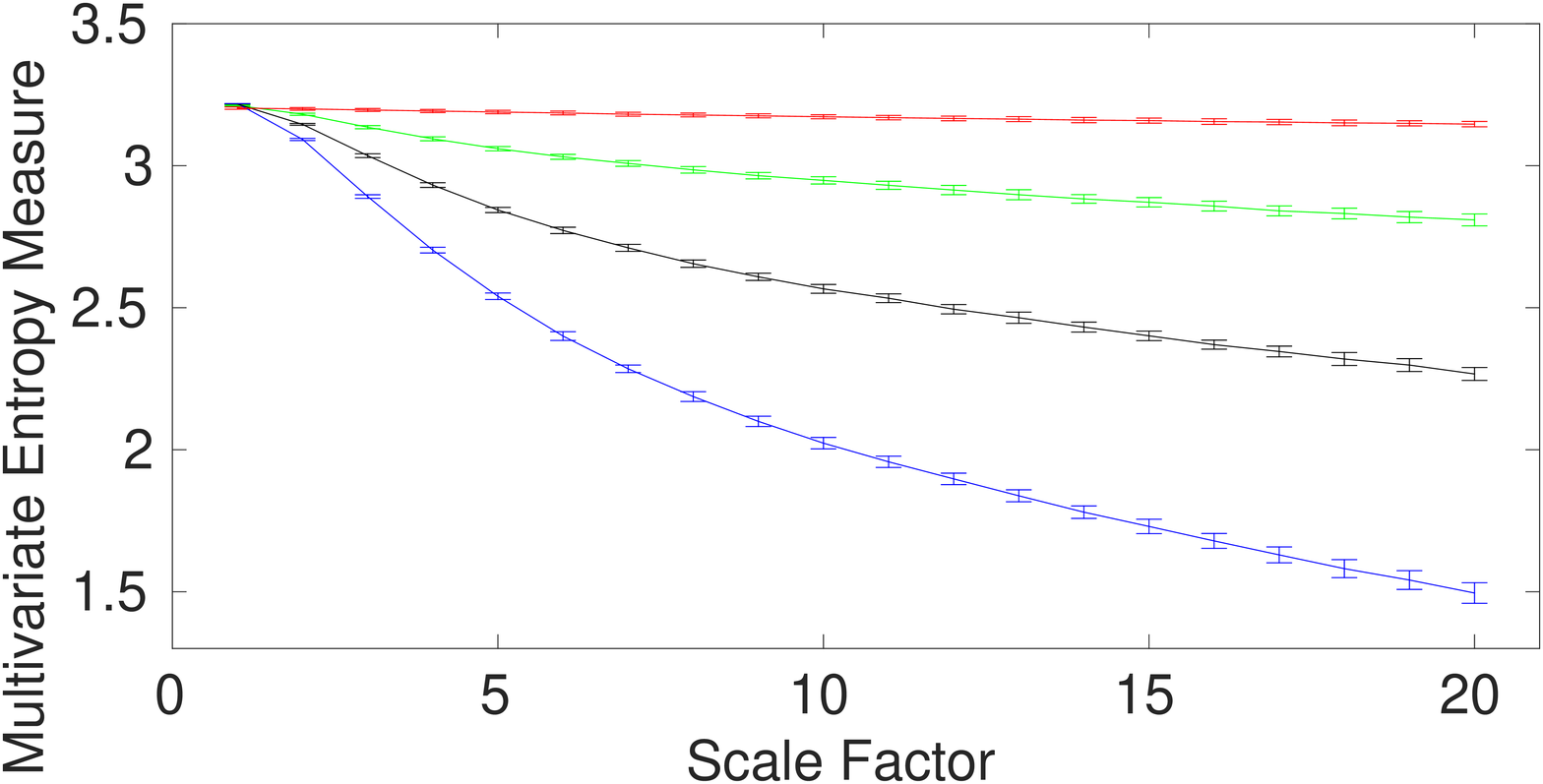}
		\footnotesize{(d) mvMDE}
		\includegraphics[width=6.2cm,height=3.5cm]{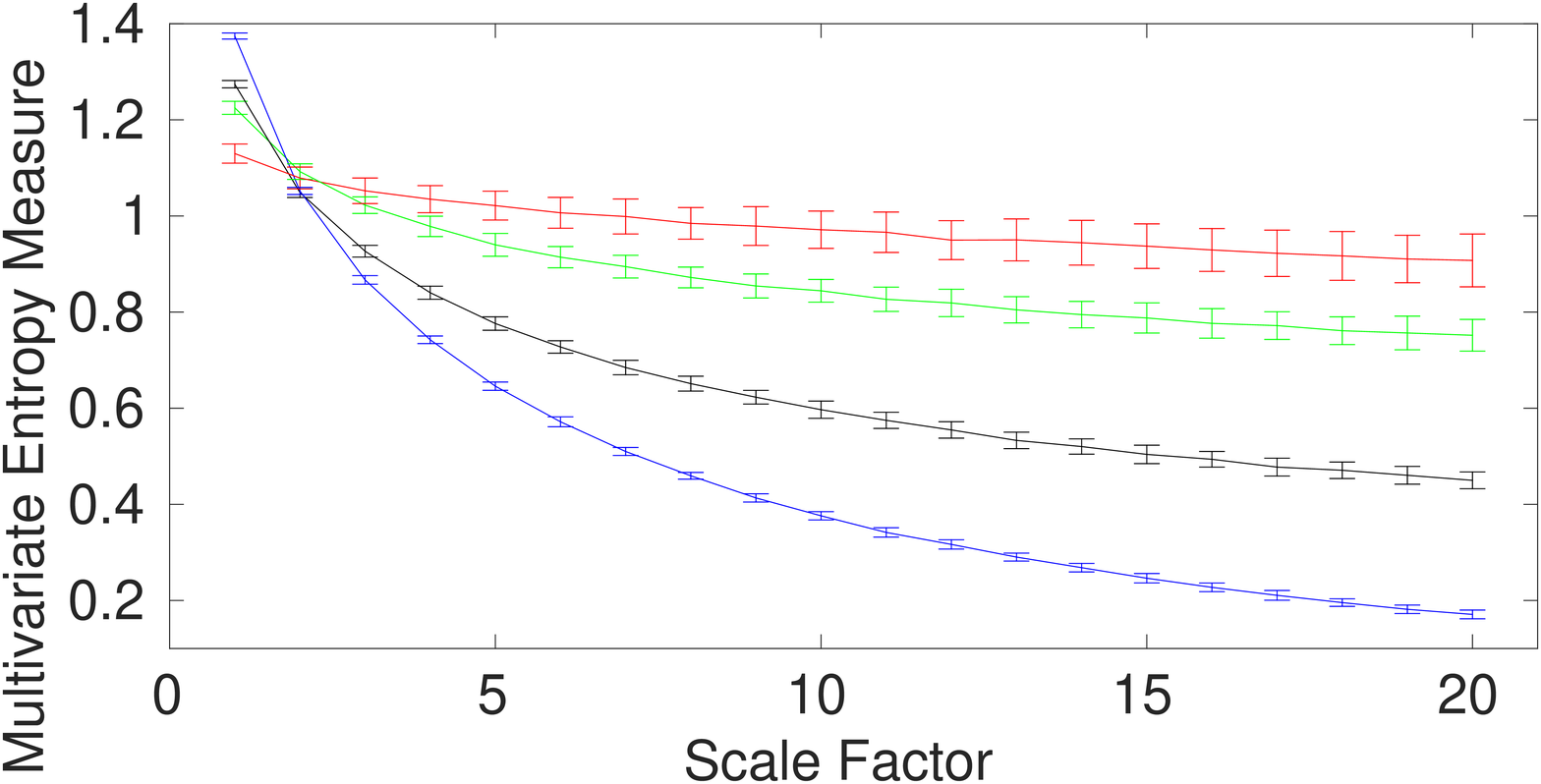}
		\footnotesize{(e) mvMSE}
		\includegraphics[width=6.2cm,height=3.5cm]{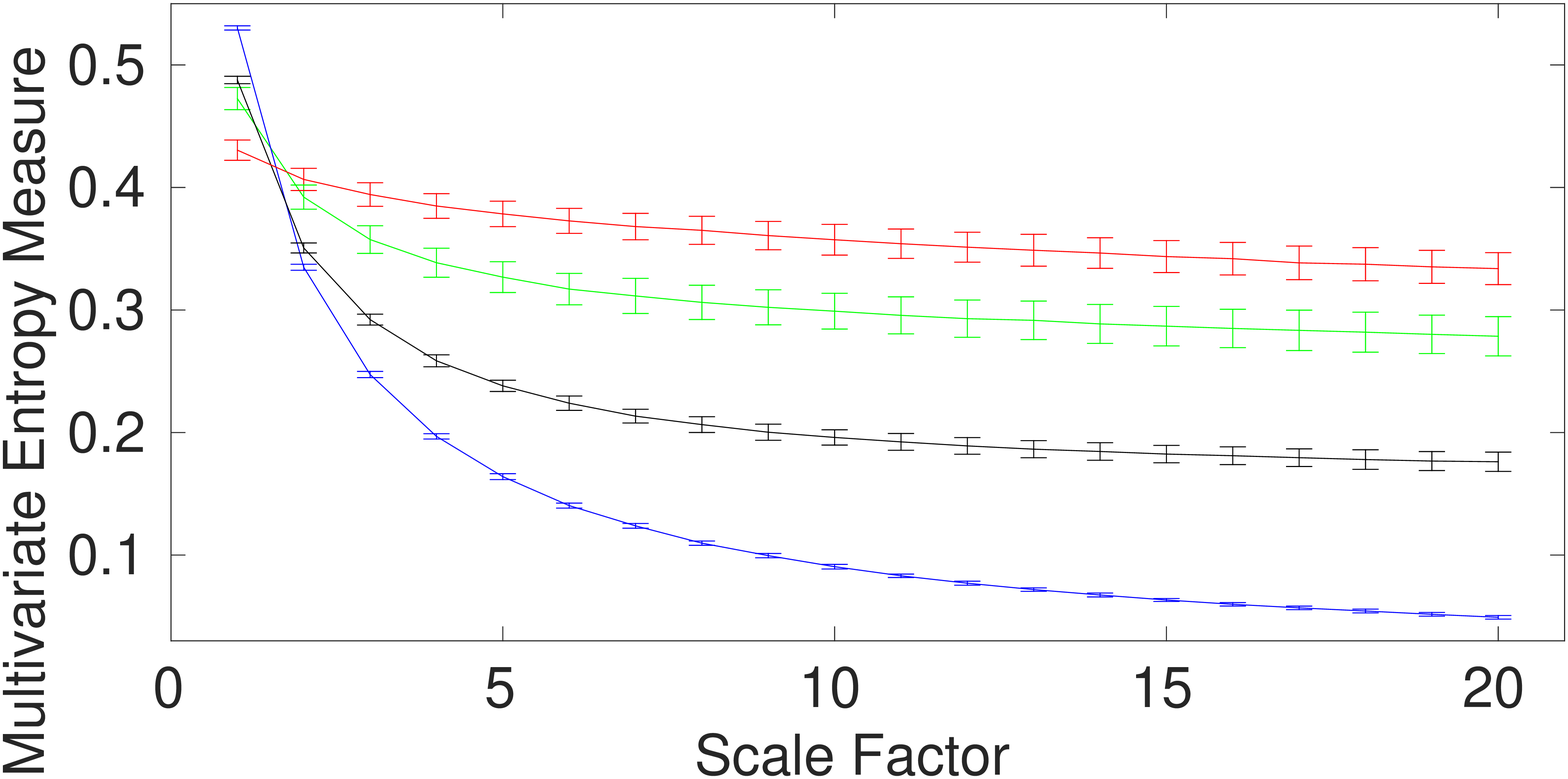}
		\footnotesize{(f) mvMFE}
		
	\end{multicols}
	
	\caption{\footnotesize{Mean value and SD of the results using (a) mvMDE-I, (b) mvMDE-II, (c) mvMDE-III, (d) mvMDE, (e) mvMSE, and (f) mvMFE computed from 40 different uncorrelated trivariate WGN and $ 1/f $ noise time series with length 15000.}}
	\label{figurelabel}
		
\end{figure*}

To compare the results obtained by the mvMDE, mvMSE, and mvMFE methods, we used the coefficient of variation (CV) defined as the SD divided by the mean. In fact, CV permits comparison of variability estimates regardless of the magnitude values. We investigate the results obtained by uncorrelated noise signals at scale factor 10, as a trade-off between short and long scale factors. As can be seen in Table II, the smallest CV values for uncorrelated trivariate $1/f$ noise, an uncorrelated combination of bivariate $1/f$ noise and univariate WGN, an uncorrelated combination of bivariate WGN and univariate $1/f$ noise, and trivariate WGN are achieved by mvMDE, mvMDE-II, mvMDE-II, and mvMDE-I, respectively. Overall, the smallest CV values for trivariate $1/f$ noise and WGN profiles are reached by the mvMDE methods, showing the superiority of the mvMDE methods over mvMSE and mvMFE in terms of stability of results.

\begin{table*}
	\centering
	\label{table_example}
	\centering     
	\caption{\footnotesize{CV values of the proposed and existing multivariate multiscale entropy-based analyses at scale factor 10 for the uncorrelated trivariate $ 1/f $ noise and WGN.}}
	\centering     
	\begin{tabular}{c*{6}{c}}
		Time series &mvMDE-I& mvMDE-II& mvMDE-III& mvMDE& mvMSE & mvMFE \\
		\hline
		All three channels contain $ 1/f$ noise                     &0.0028   & 0.0025 & 0.0037  & 0.0022& 0.0405 & 0.0355 \\
		Two channels contain $ 1/f $ noise and one contains WGN     &0.0042   &0.0032 &0.0036 &0.0044&0.0283  &0.0274  \\
		One channel contains $ 1/f $ noise and two contain WGN      & 0.0066 & 0.0052 &0.0058& 0.0061& 0.0305 & 0.0292 \\
		All three channels contain WGN                              &0.0072   & 0.0080& 0.0092 & 0.0101  &0.0232  &0.0211
	\end{tabular}
\end{table*}

To assess the ability of the mvMDE methods to characterize short signals in comparison with mvMFE and mvMSE, we use trivariate $ 1/f $ and WGN noise with length of 300 sample points. The results for the mvMDE, mvMSE, and mvMFE approaches at temporal scales 1 to 20 are depicted in Fig. 2(a) to 2(f), respectively. As can be seen in Fig. 2(a) and 2(d), the mvMDE-I and mvMDE methods better discriminate different dynamics of the noise signals. However, the mvMSE values are undefined at higher scale factors. Although the mvMFE- and mvMDE-II-based values are defined at all scale factors, they cannot distinguish the dynamics of different noise signals. The profiles obtained by mvMDE-III are more distinguishable than mvMDE-II, as mentioned that mvMDE-III needs a smaller number of sample points. Nevertheless, the profiles obtained by mvMDE-III have overlaps at several scale factors. Overall, the results show the superiority of mvMDE-I and mvMDE over mvMDE-II, mvMDE-III, mvMSE, and mvMFE for short uncorrelated signals.

\begin{figure*}
	\centering
	\begin{multicols}{3}
		\includegraphics[width=6.2cm,height=3.5cm]{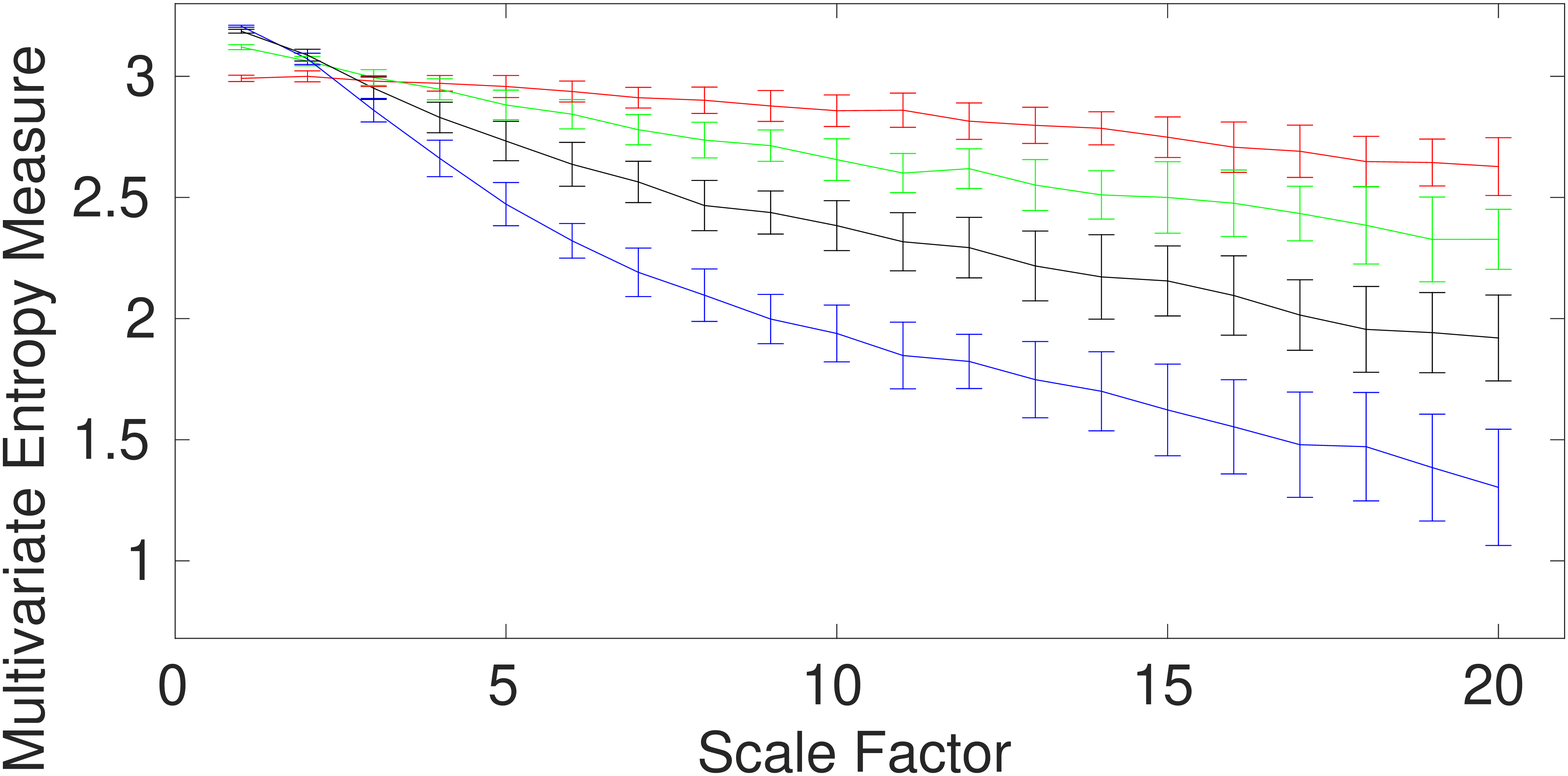}
		\footnotesize{(a) mvMDE-I}
		\includegraphics[width=6.2cm,height=3.5cm]{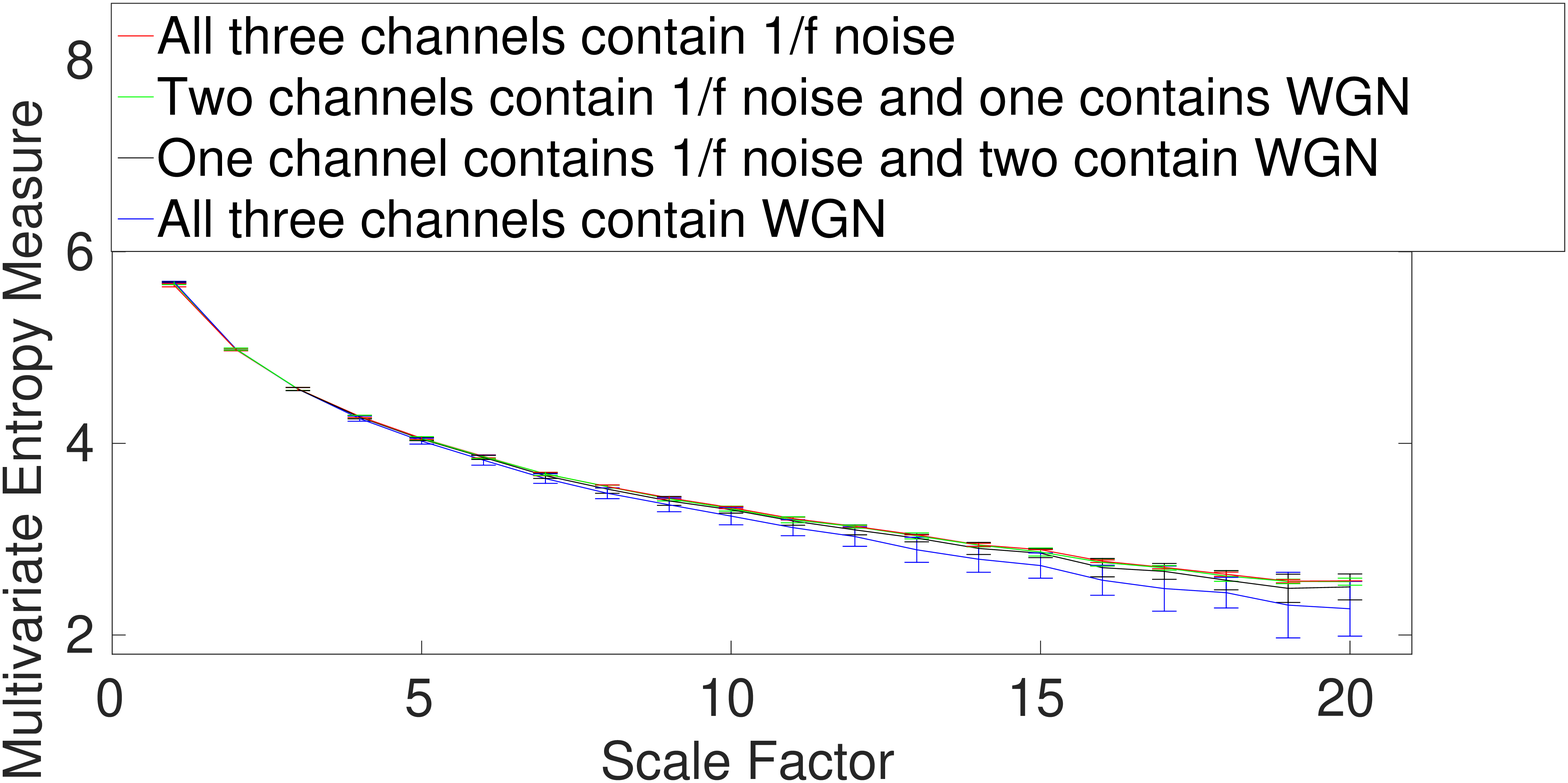}
		\footnotesize{(b) mvMDE-II}
		\includegraphics[width=6.2cm,height=3.5cm]{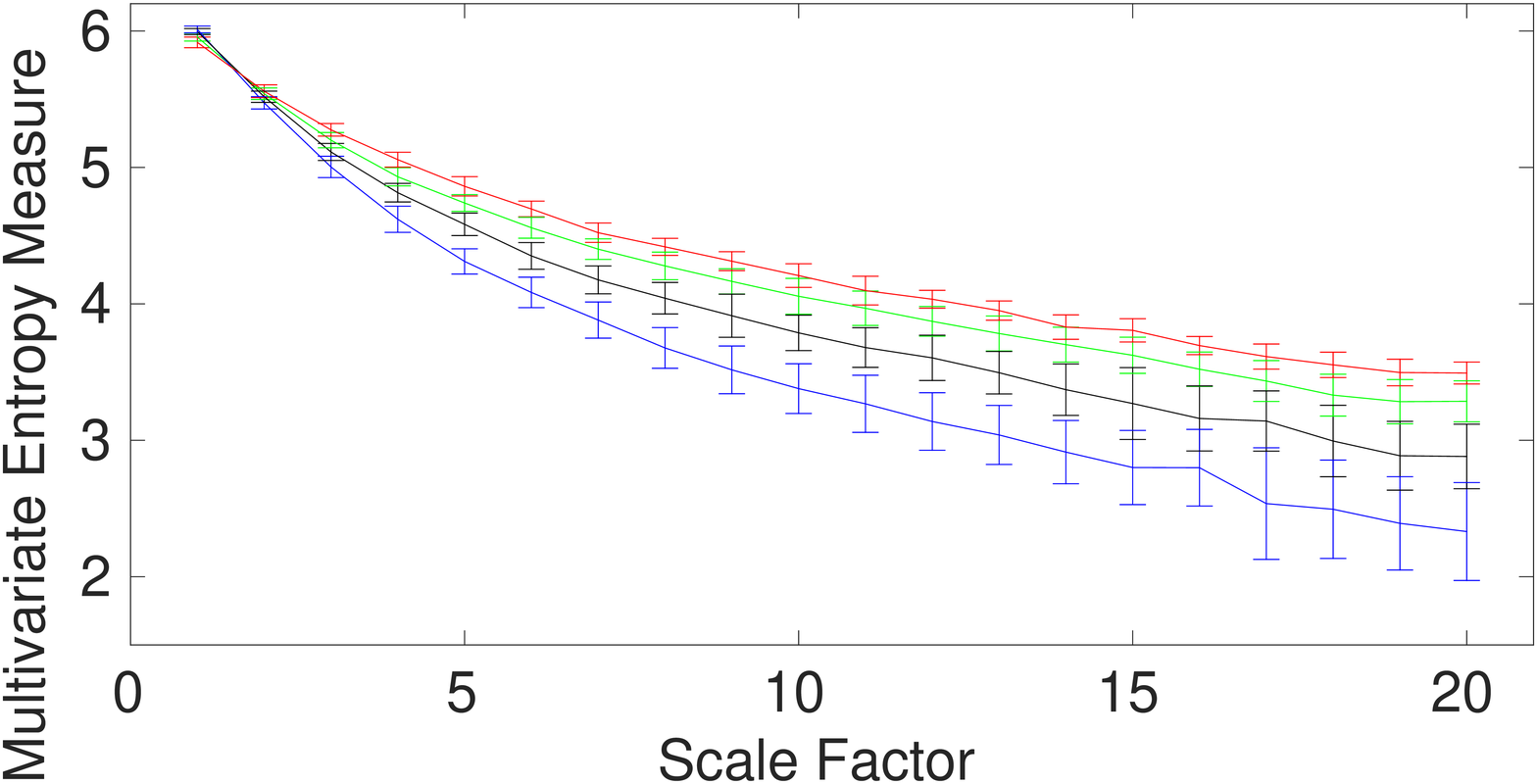}
		\footnotesize{(c) mvMDE-III}
	\end{multicols}
	\begin{multicols}{3}
		\centering
		\includegraphics[width=6.2cm,height=3.5cm]{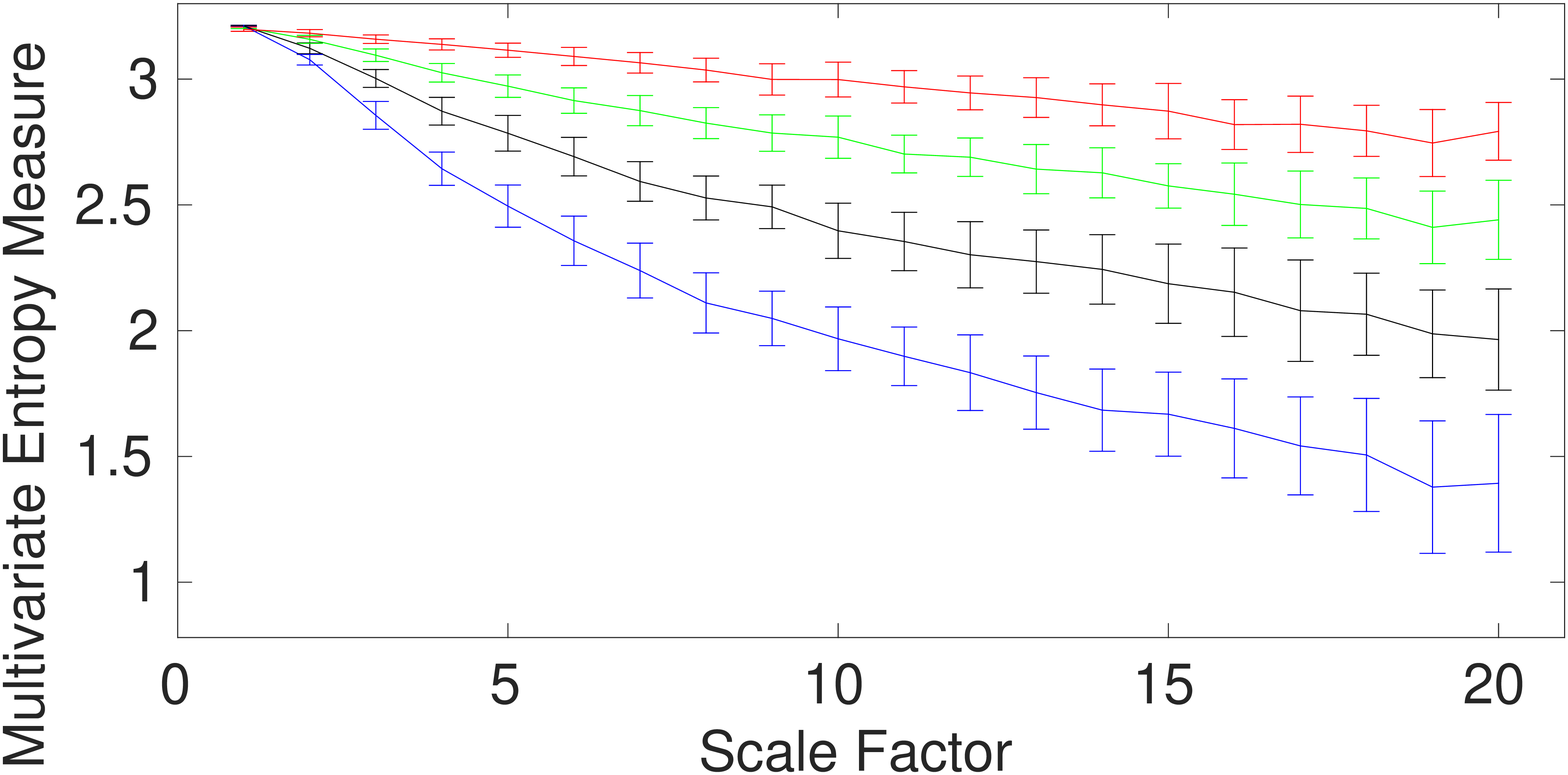}
		\footnotesize{(d) mvMDE}
		\includegraphics[width=6.2cm,height=3.5cm]{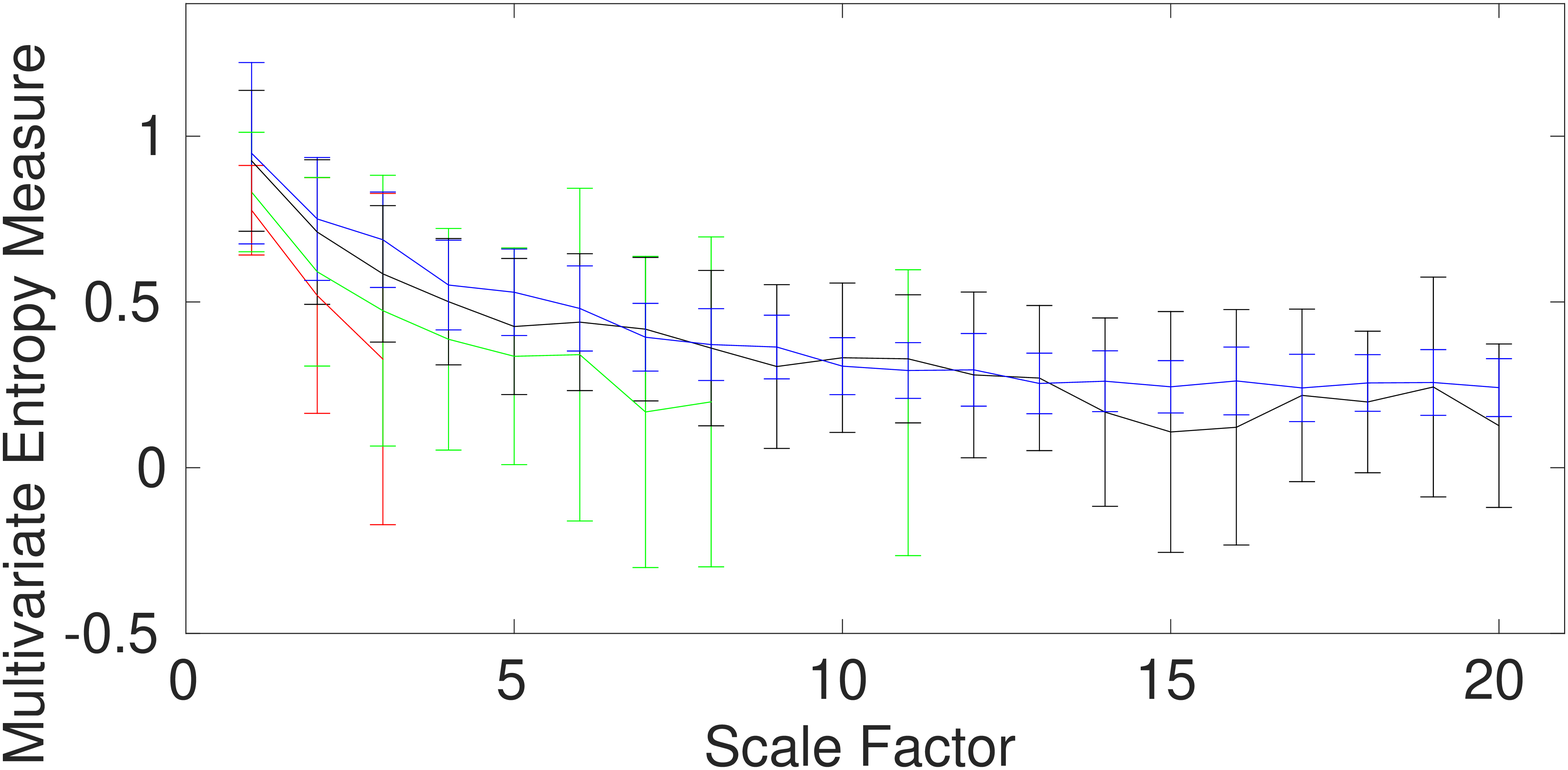}
		\footnotesize{(e) mvMSE}
		\includegraphics[width=6.2cm,height=3.5cm]{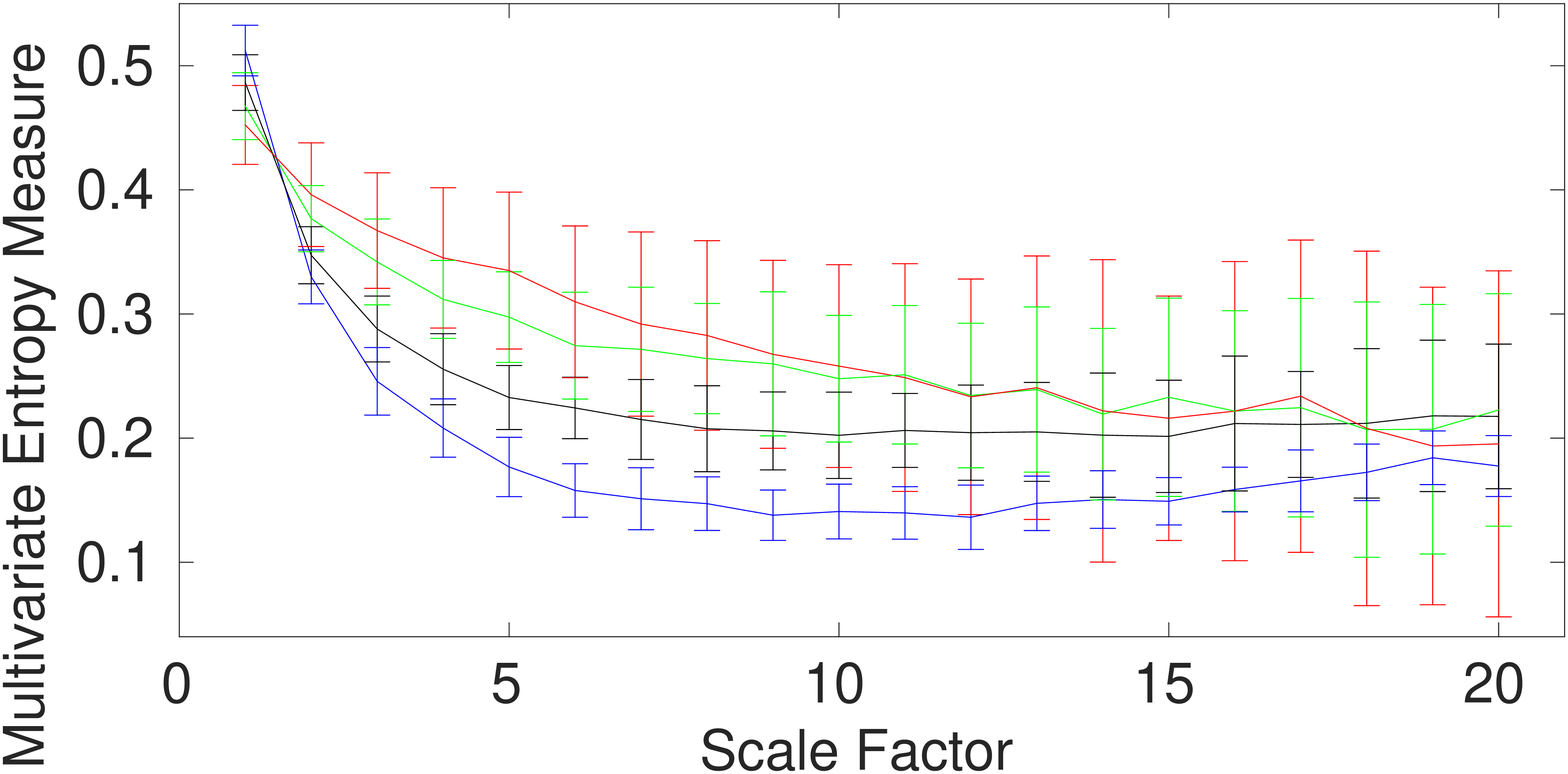}
		\footnotesize{(f) mvMFE}
		
	\end{multicols}
	
	\caption{\footnotesize{Mean value and SD of the results obtained by (a) mvMDE-I, (b) mvMDE-II, (c) mvMDE-III, (d) mvMDE, (e) mvMSE, and (f) mvMFE computed from 40 different uncorrelated trivariate WGN and $ 1/f $ noise time series with length 300.}}
	\label{figurelabel}
\end{figure*}

To evaluate the computational time of mvMSE, mvMFE, mvMDE-I to III, and mvMDE, we use uncorrelated multivariate WGN time series with different lengths, changing from 100 to 10,000 sample points, and different number of channels, changing from 2 to 8. The results are depicted in Table III. The simulations have been carried out using a PC with Intel (R) Xeon (R) CPU, E5420, 2.5 GHz and 8-GB RAM by MATLAB R2015a. The results show that the computation times for mvMSE and mvMFE are close. The slowest algorithm is mvMDE-II, while the fastest ones are mvMDE-I and mvMDE, in that order. For an 8-channel signal with 10,000 samples, using mvMSE, mvMFE, and mvMDE-II, the array exceeded the memory available. Overall, in terms of computation time and memory space, mvMDE outperforms all the existing and proposed methods taking into account both the time and spatial domains.  

\begin{table*}
	\centering
	\label{tab:table4}\caption{Computational time of the mvMSE, mvMFE, and mvMDE algorithms with $\tau_{max}=10$.} 
	\begin{tabular}{c*{8}{c}}
		Number of channels and samples   &mvMSE& mvMFE& mvMDE-I& mvMDE-II& mvMDE-III& mvMDE \\
		\hline
		2 channels and 1,000 samples  & 0.141 s& 0.153 s& 0.083 s  & 0.116 s  &0.100 s &0.089 s  \\
		2 channels and 3,000 samples  & 0.598 s & 0.723 s&0.240 s & 0.3126 s  & 0.280 s  & 0.265 s \\
		2 channels and 10,000 samples & 4.234 s & 5.334 s&0.736  s   & 1.010 s  & 0.919 s & 0.868 s \\
		5 channels and 1,000 samples  & 0.544 s & 0.636 s& 0.191 s& 91.240 s&0.903 s  & 0.229 s \\
		5 channels and 3,000 samples  & 3.174 s & 3.586 s& 0.568 s &169.275 s&2.209 s  & 0.670 s \\
		5 channels and 10,000 samples & 28.229 s &31.242 s& 1.850 s  &454.199 s&7.271 s& 2.312 s  \\			
		8 channels and 1,000 samples  & 1.479 s &1.573 s&  0.298 s &out of memory error& 103.096 s&0.354 s\\
		8 channels and 3,000 samples  & 9.421 s & 9.972 s& 0.820 s&out of memory error&245.034 s  &1.028 s \\
		8 channels and 10,000 samples &out of memory error&out of memory error&2.687 s&out of memory error& 745.633 s&3.509 s
		
	\end{tabular}
\end{table*}

Univariate multiscale entropy approaches only consider every data channel separately and fail to take into account the cross-channel information of multivariate time series \cite{ahmed2011multivariate}. Uncorrelated multi-channel WGN has less structural complexity and more irregularity compared with multi-channel $1/f$ noise. To assess the ability of the existing and proposed multivariate entropy methods to reveal the dynamics across the channels, we created 40 independent realizations of different combinations of bivariate $ 1/f $ noise and WGN time series with length 20,000 (according to \cite{ahmed2011multivariate,azami2017refined}), making the channels correlated. Fig. 3(a) to 3(d) respectively show the results obtained using the mvMDE-I, mvMDE-II, mvMDE-III, and mvMDE to model both the within- and cross-channel properties in multivariate signals.

mvMDE-I cannot discriminate the correlated from uncorrelated WGN or $1/f$ noise. This fact is revealed in Fig. 3 (a). Therefore, mvMDE-I should only be used when the components of a multi-channel time series are statistically independent. Multivariate multiscale entropy-based methods at scale factor 1 show the irregularity of multi-channel signals \cite{ahmed2011multivariate}. The mvMDE-II, mvMDE-III, and mvMDE values at scale 1 show that the uncorrelated WGN is the most irregular and unpredictable time series in agreement with \cite{costa2005multiscale}, while the most irregular signals using mvMFE and mvMSE are the correlated WGN \cite{azami2017refined,ahmed2011multivariate}, in contrast with the fact that correlated multi-channel WGN signals are more predictable and regular than uncorrelated WGN ones \cite{costa2005multiscale,azami2017refineddd}.

The correlated bivariate $1/f$ noise is the most complex signal using the mvMDE-II, mvMDE-III, and mvMDE. The second most complex signal is the uncorrelated bivariate $1/f$ noise, as can be seen in Fig. 3. The decreases of the uncorrelated bivariate WGN noise profiles using mvMDE-II, mvMDE-III, and mvMDE are the largest, evidencing the fact that the uncorrelated WGN is the least complex time series. These facts are also in agreement with the previous studies \cite{ahmed2011multivariate,azami2017refined,fogedby1992phase}. Therefore, as desired, the mvMDE-II, mvMDE-III, and mvMDE deal with both the cross- and within-channel correlations.

\begin{figure*}
	\centering
	\begin{multicols}{4}
		\includegraphics[width=4.6cm,height=3.5cm]{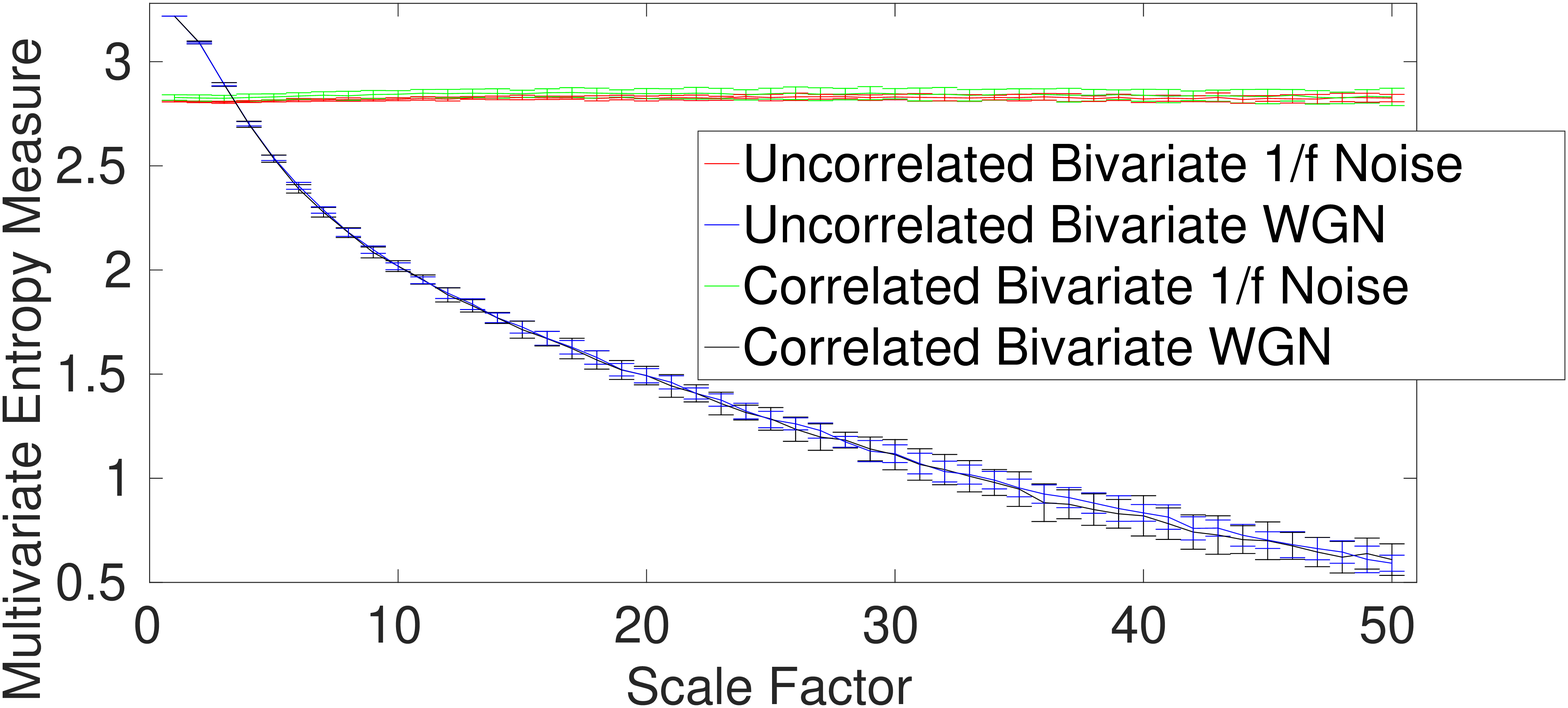}
		\footnotesize{(a) mvMDE-I}
		\includegraphics[width=4.6cm,height=3.5cm]{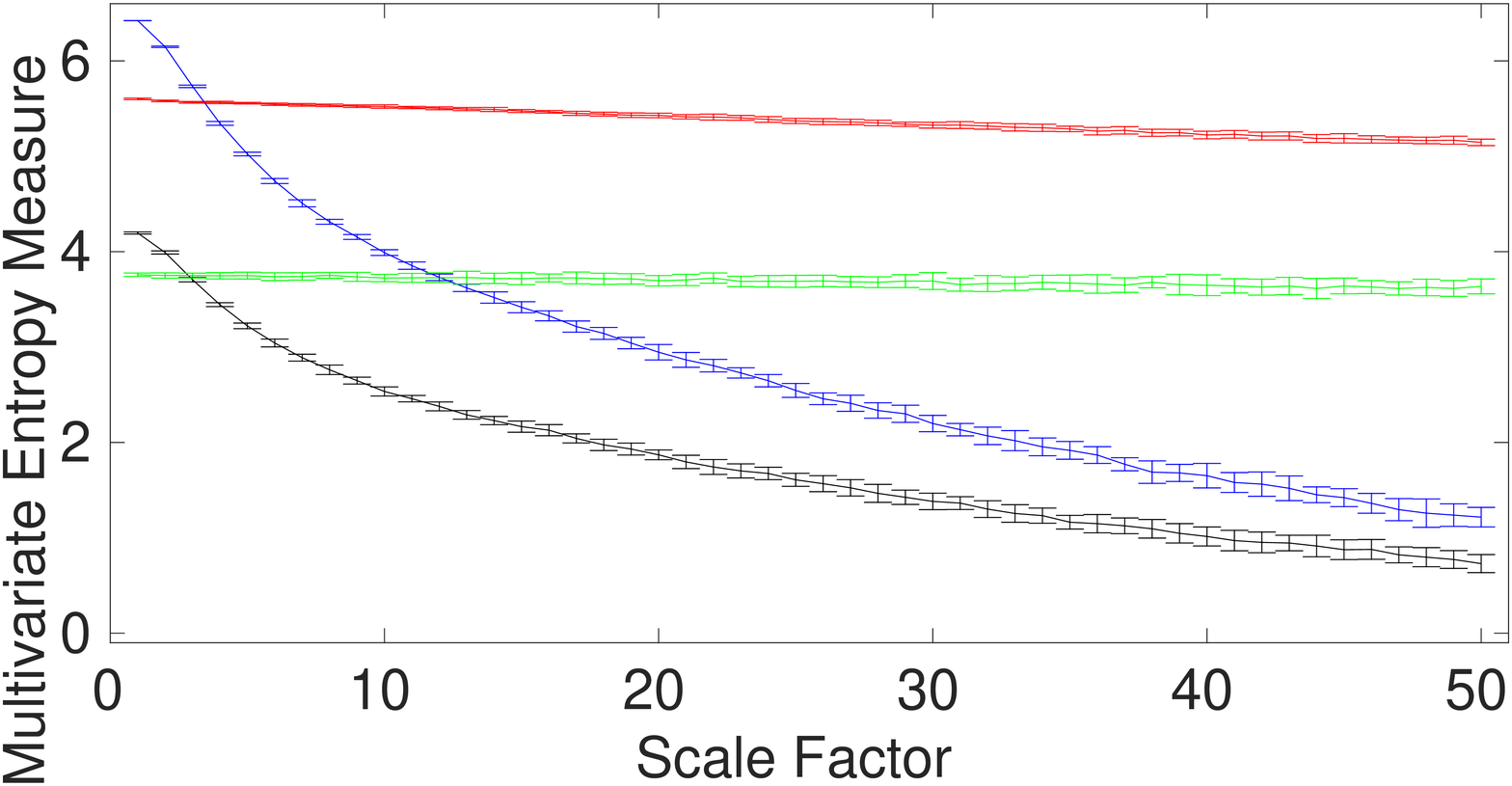}
		\footnotesize{(b) mvMDE-II}
		\includegraphics[width=4.6cm,height=3.5cm]{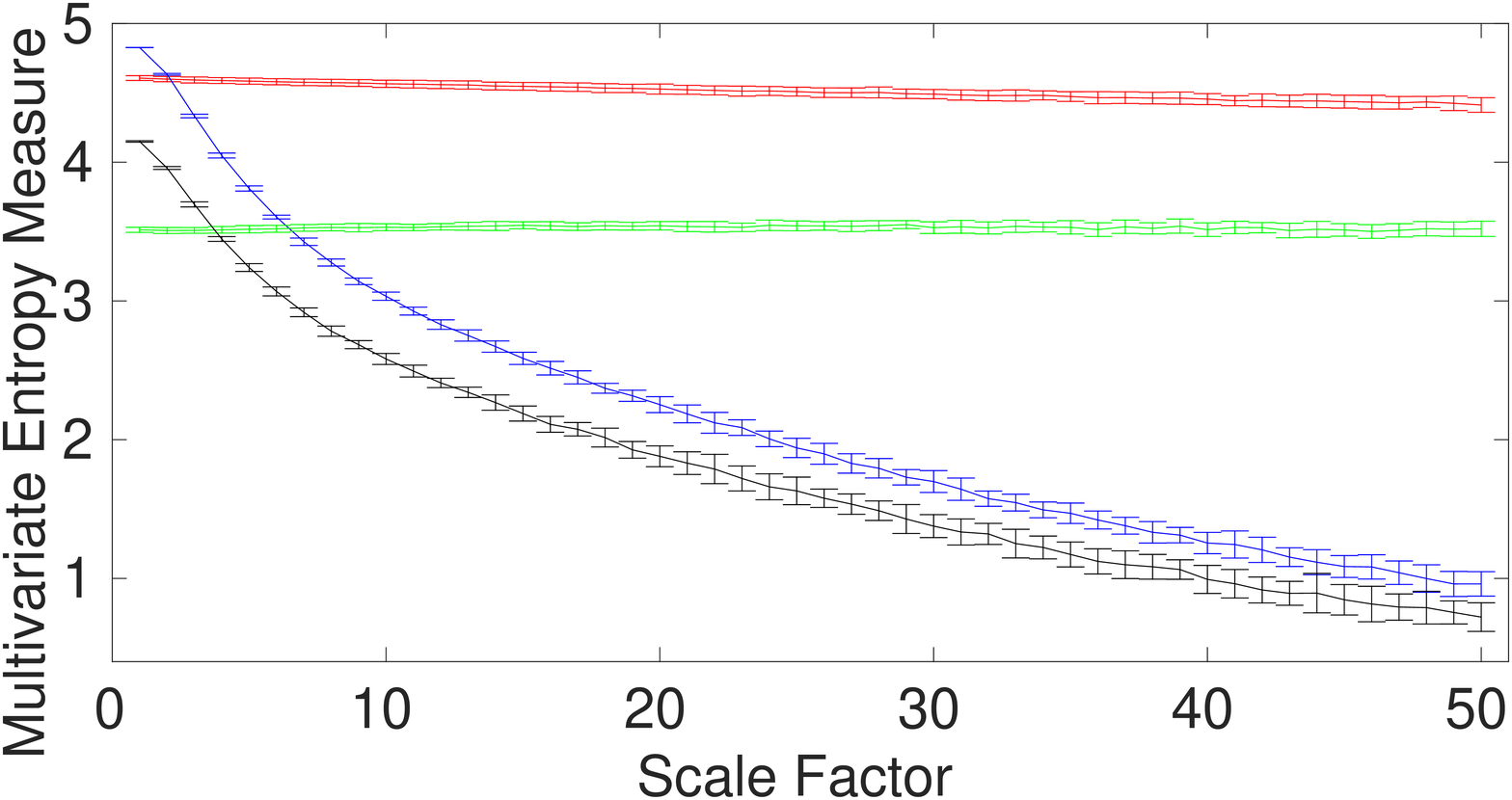}
		\footnotesize{(c) mvMDE-III}
	\includegraphics[width=4.6cm,height=3.5cm]{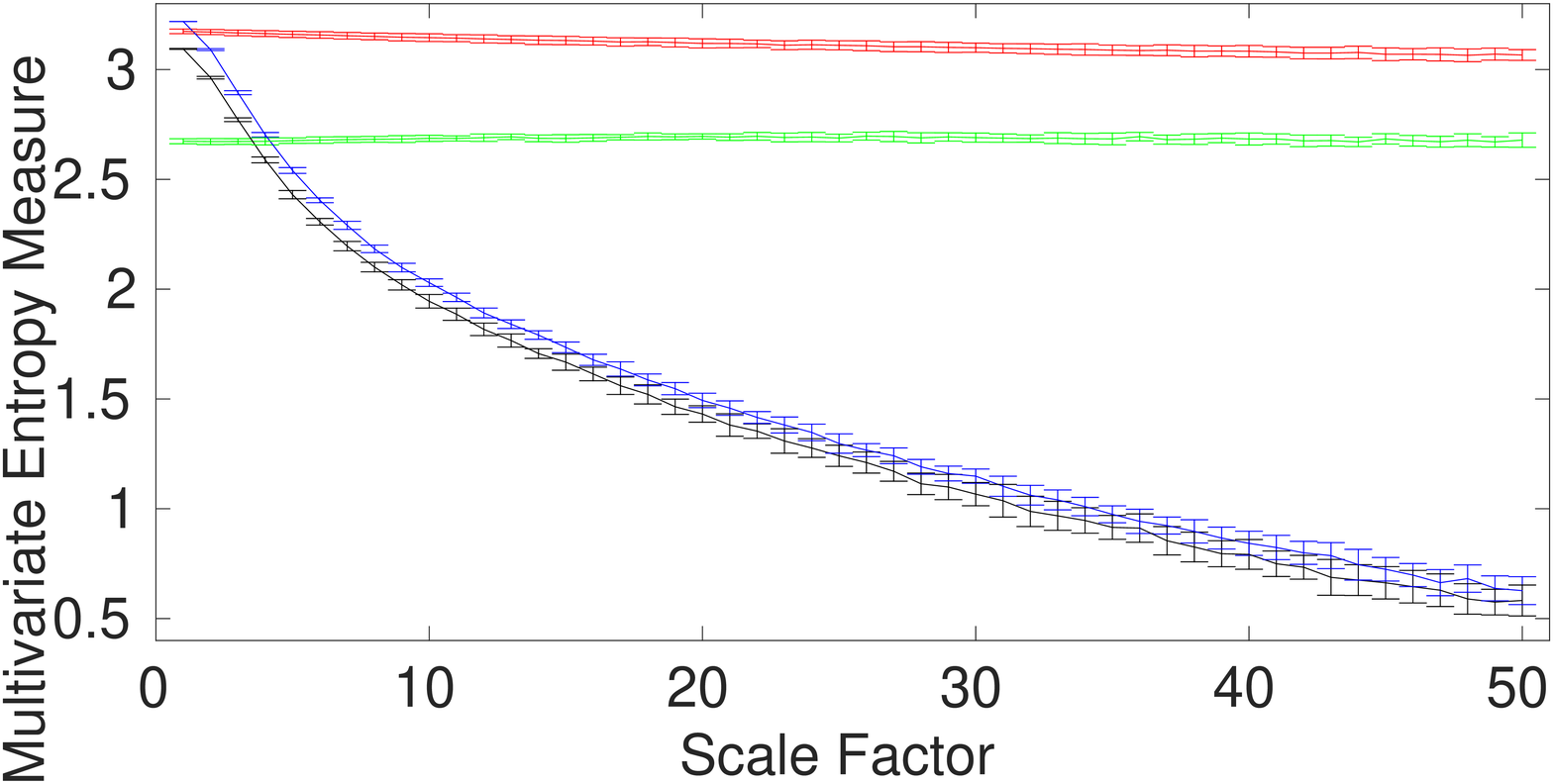}
	\footnotesize{(d) mvMDE}
	\end{multicols}
		
	\caption{\footnotesize{Mean value and SD of the results obtained by (a) mvMDE-I, (b) mvMDE-II, (c) mvMDE-III, and (d) mvMDE computed from 40 different correlated and uncorrelated bivariate WGN and $ 1/f $ noise time series with length 300.}}
	\label{figurelabel}

\end{figure*}

The ability of the mvMDE methods to characterize multivariate AR processes is further evaluated using bivariate AR(1), AR(3), and AR(5). The results obtained by the mvMDE-I, mvMDE-II, mvMDE-III, and mvMDE are shown in Fig. 4(a) to 4(d). As expected, when the lag order increases, the complexity of the corresponding time series using the mvMDE approaches increases, in agreement with the fact that a larger lag order denotes a more complex time series \cite{ahmed2011multivariate}.

\begin{figure*}
	\centering
	\begin{multicols}{4}
		\includegraphics[width=4.6cm,height=3.5cm]{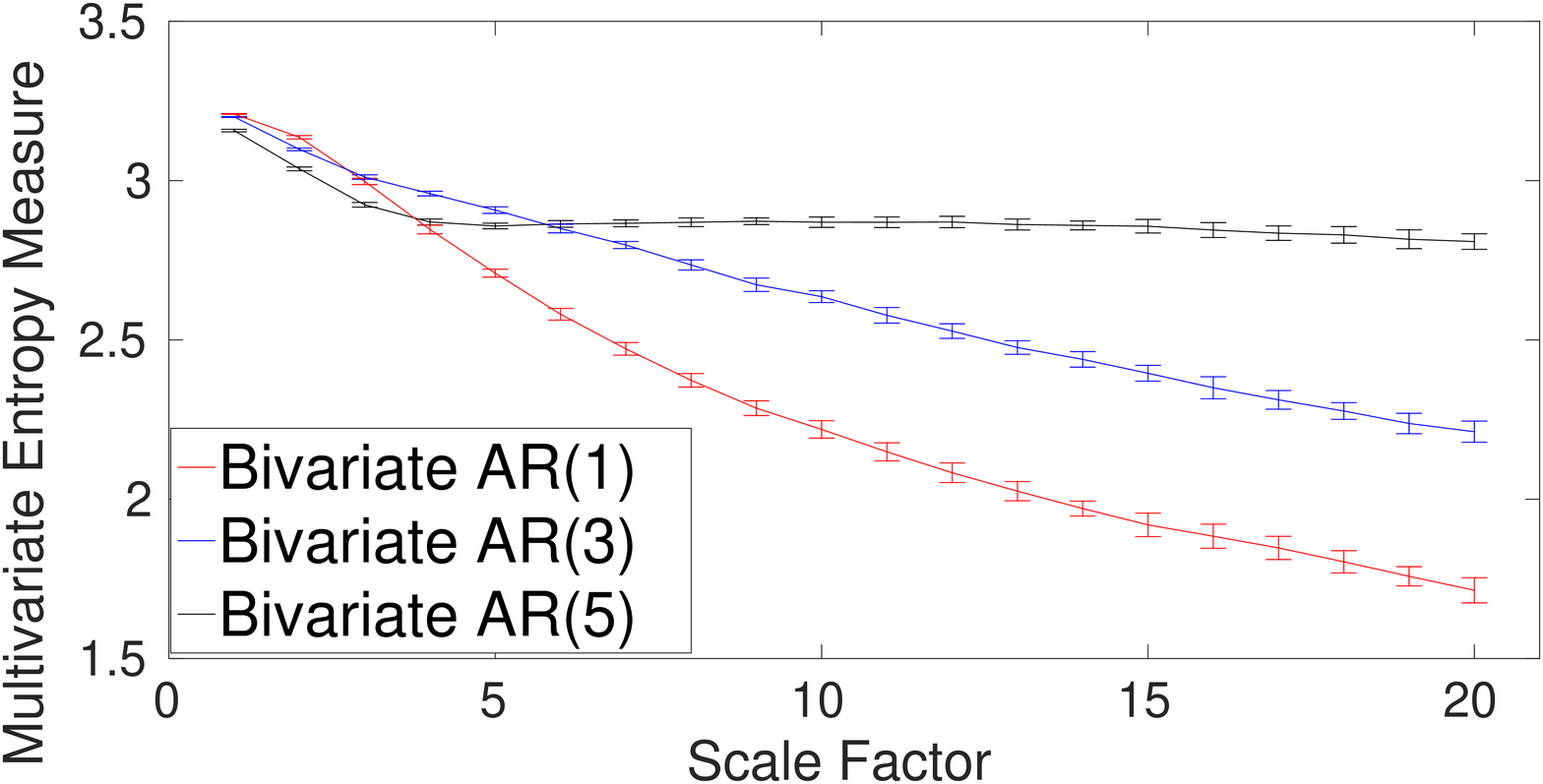}
		\footnotesize{(a) mvMDE-I}
		\includegraphics[width=4.6cm,height=3.5cm]{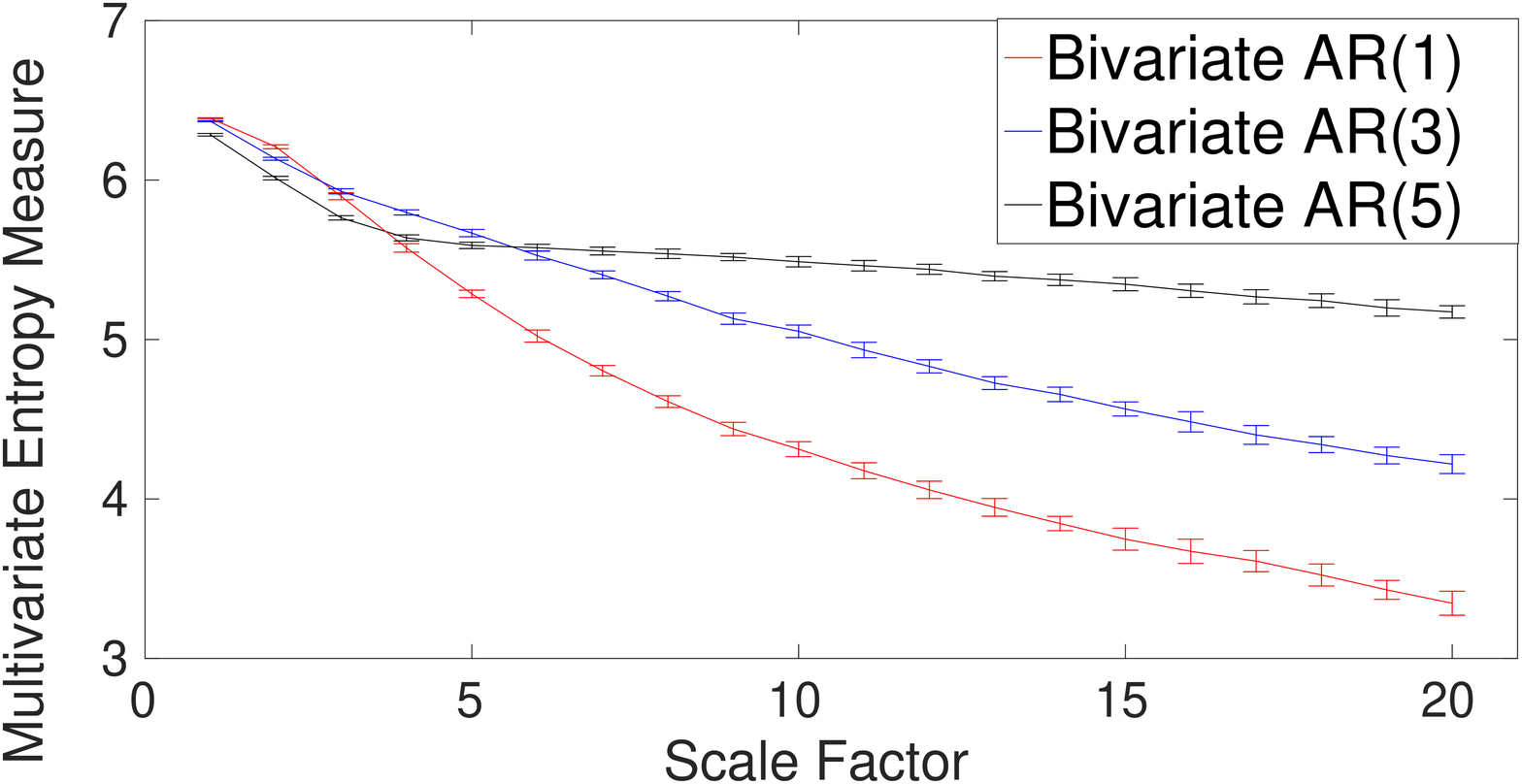}
		\footnotesize{(b) mvMDE-II}
		\includegraphics[width=4.6cm,height=3.5cm]{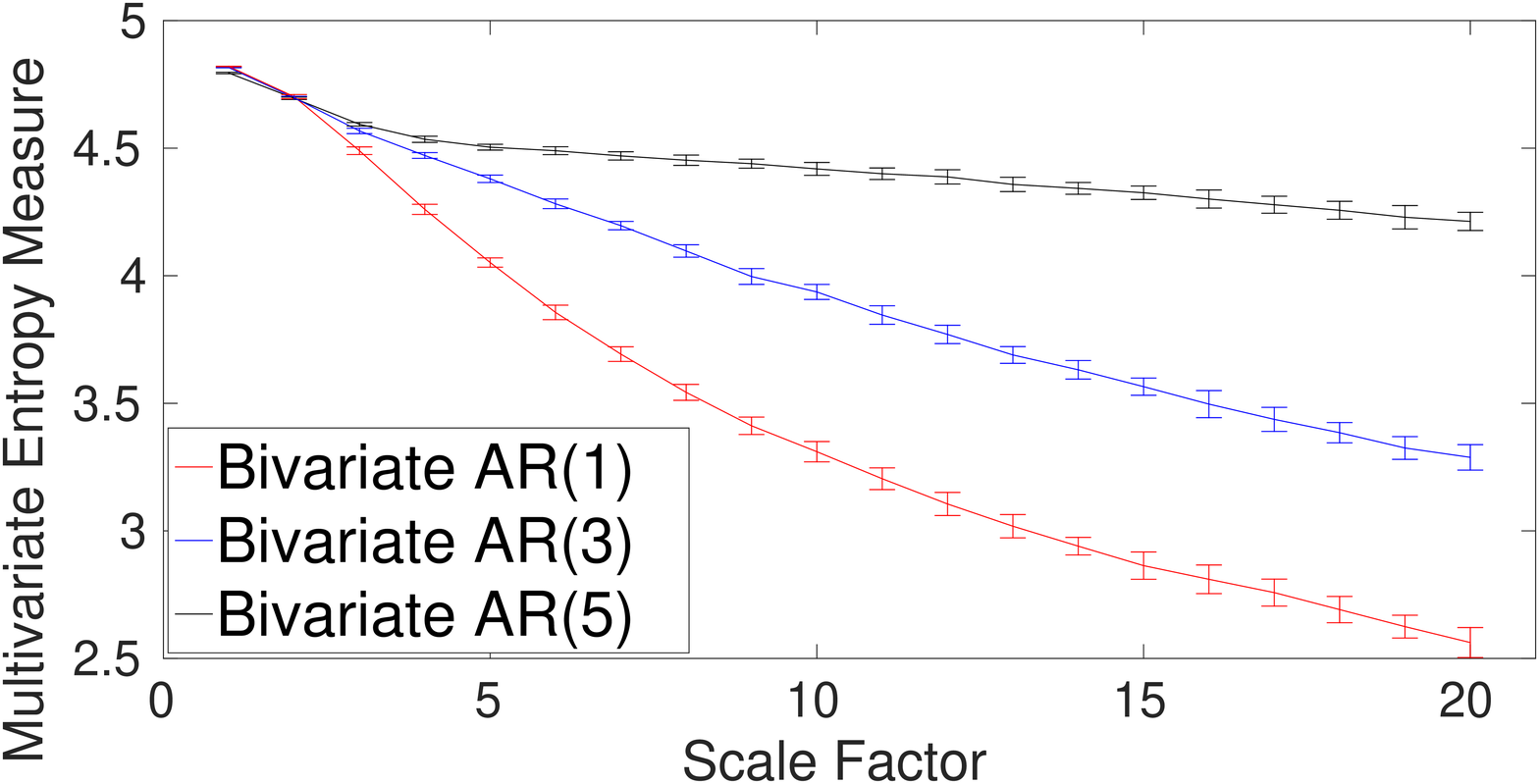}
		\footnotesize{(c) mvMDE-III}
		\includegraphics[width=4.6cm,height=3.5cm]{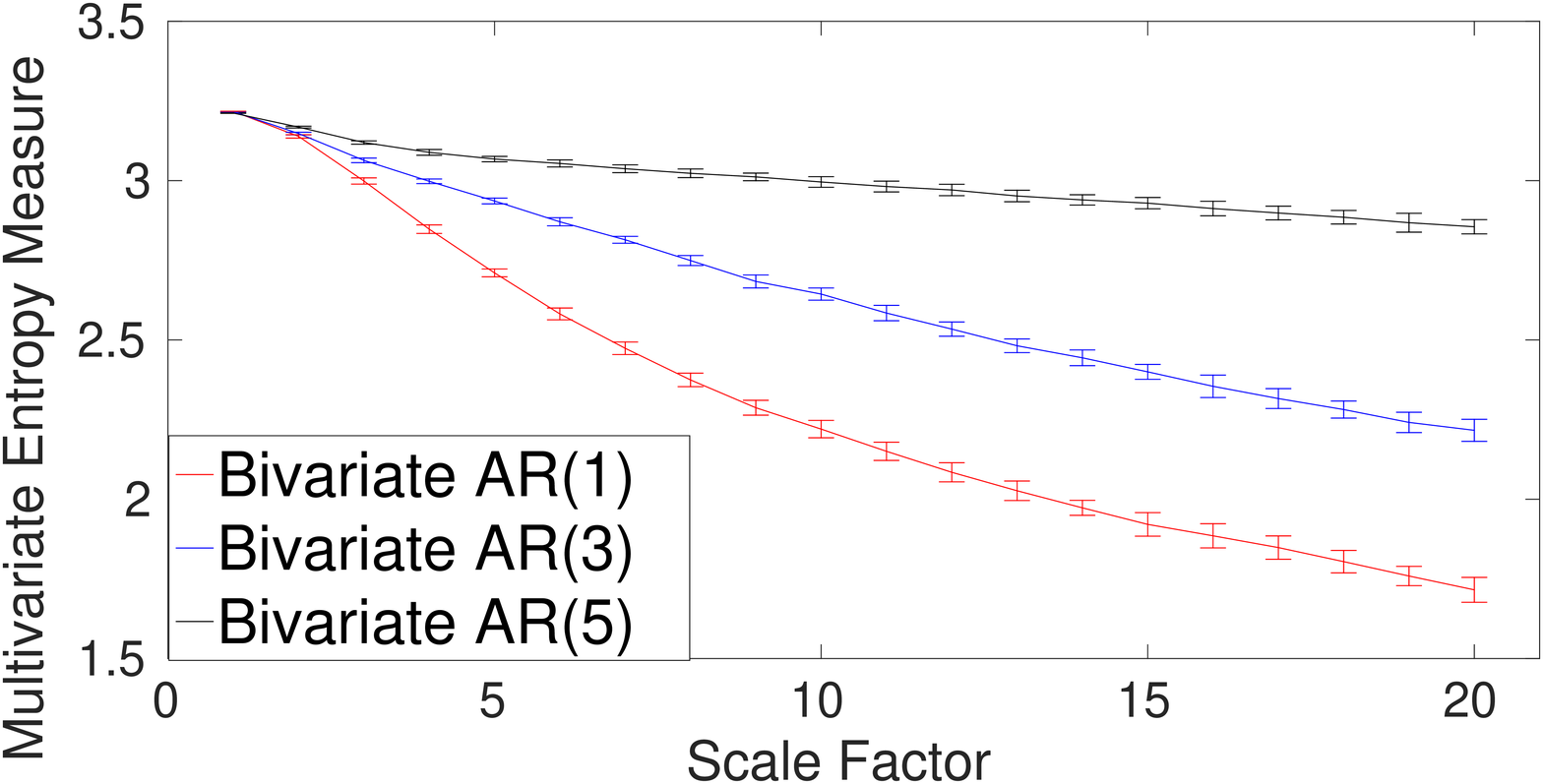}
		\footnotesize{(d) mvMDE}
	\end{multicols}
	
	\caption{\footnotesize{Average and SD values of the results using (a) mvMDE-I, (b) mvMDE-II, (c) mvMDE-III, and (d) mvMDE computed from 40 different bivariate AR(1), AR(3), and AR(5) time series with length 10000.}}
	\label{figurelabel}

\end{figure*}

\subsection{Real Biomedical Datasets}
Discrimination of aged and diseased individuals' from control or healthy subjects' time series is a long-lasting challenge in the physiological complexity literature \cite{costa2005multiscale,azami2017refined,ahmed2011multivariate,Labate2013}. To this end, we use the mvMDE methods, in comparison with mvMFE as an improved version of mvMSE \cite{azami2017refined}, to detect different types of dynamical variability of multivariate recordings of three physiological datasets. Of note is that we do not use the mvMDE-I for biomedical signals, because it does not take into account both the spatial and time domains at the same time.

\textit{1) Dataset of Stride Internal Fluctuations}: For the self-paced versus metronomically-paced stride interval fluctuations, the results obtained by the mvMDE-III, mvMDE, and mvMFE, respectively depicted in Fig. 5(a), (b), and (c), show that the self-paced unconstrained walk's fluctuations have more complexity and greater long-range correlations than the metronomically-paced walk's series, in agreement with those obtained by mvMSE, and multivariate empirical mode decomposition-enhanced by mvSE \cite{ahmed2012dynamical}. We did not use mvMDE-II, as the signals do not follow the minimum number of samples required for mvMDE-II.
To compare the results, the CV values for both the metronomically- and self-paced walk (MPW and SPW) at scale factor 4, as a trade-off between the long and short scales, are shown in Table IV. The CV values for the mvMDE-III- and mvMDE-based profiles are smaller than those for mvMFE, showing the superiority of the proposed methods over mvMFE in terms of the stability of results. The smallest CV values are achieved by the mvMDE.

\begin{figure*}
	\centering
	\begin{multicols}{3}
		\includegraphics[width=6.2cm,height=3.5cm]{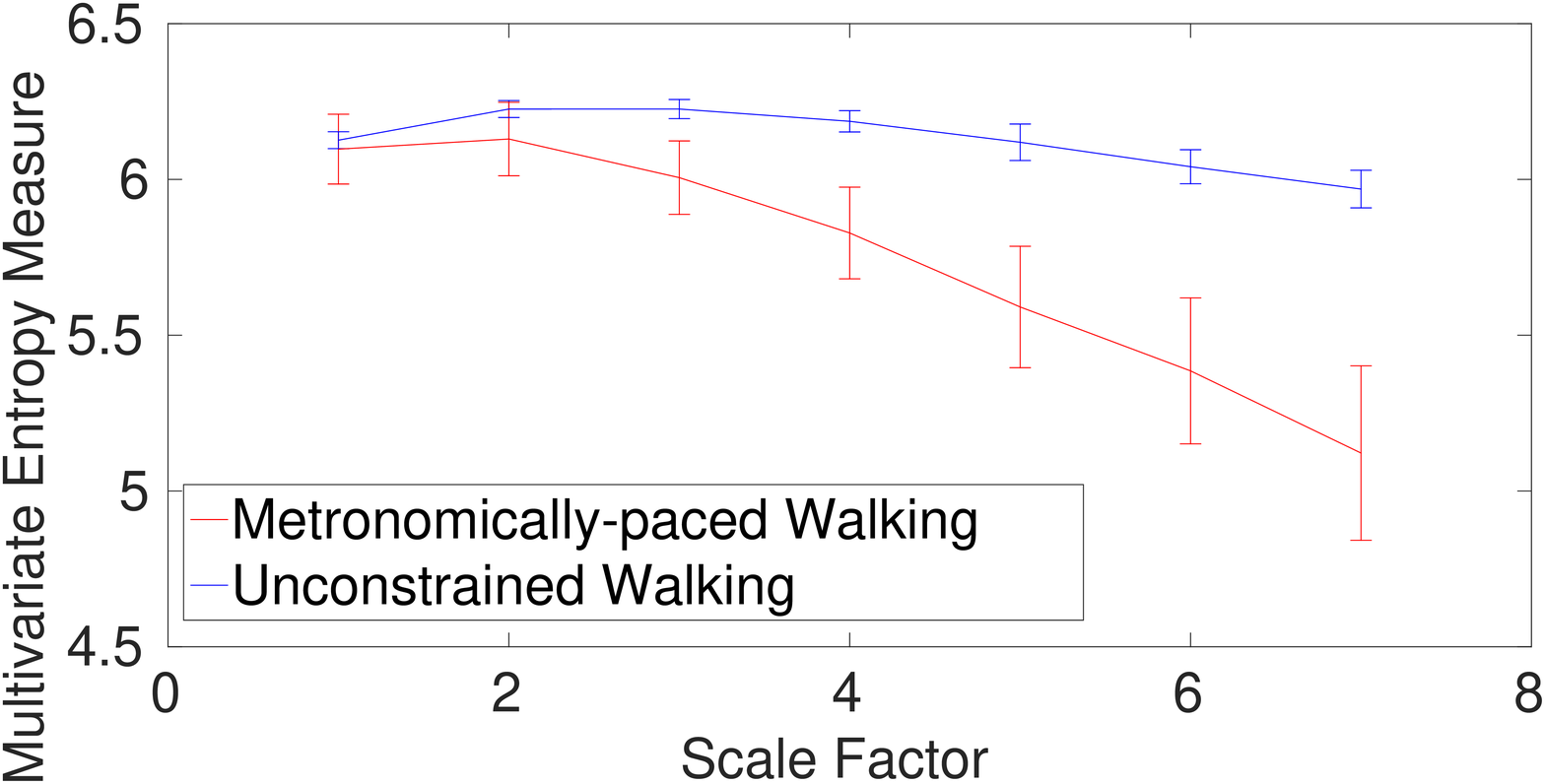}
		\footnotesize{(a) mvMDE-III}
		\includegraphics[width=6.2cm,height=3.5cm]{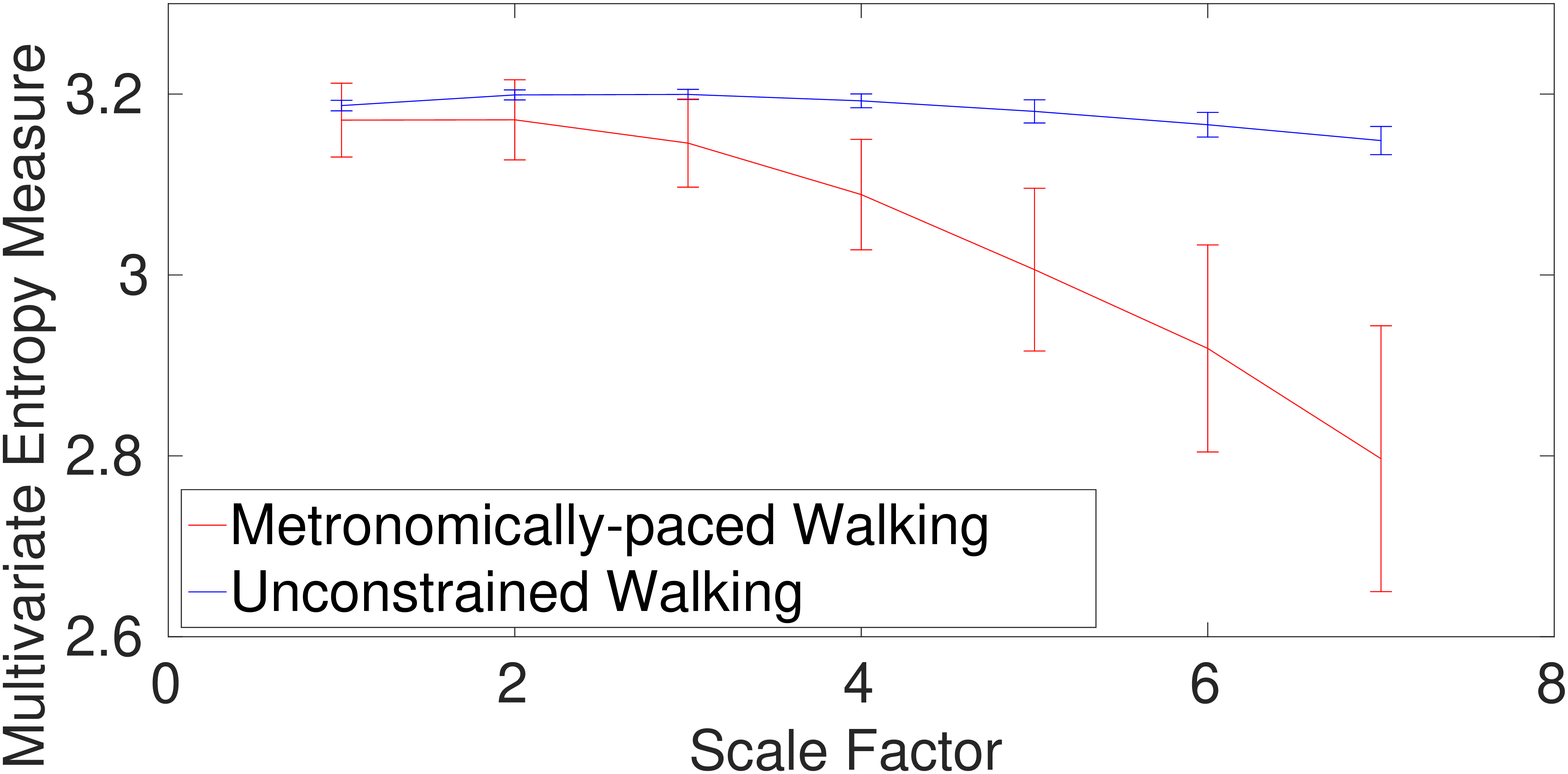}
		\footnotesize{(b) mvMDE}
		\includegraphics[width=6.2cm,height=3.5cm]{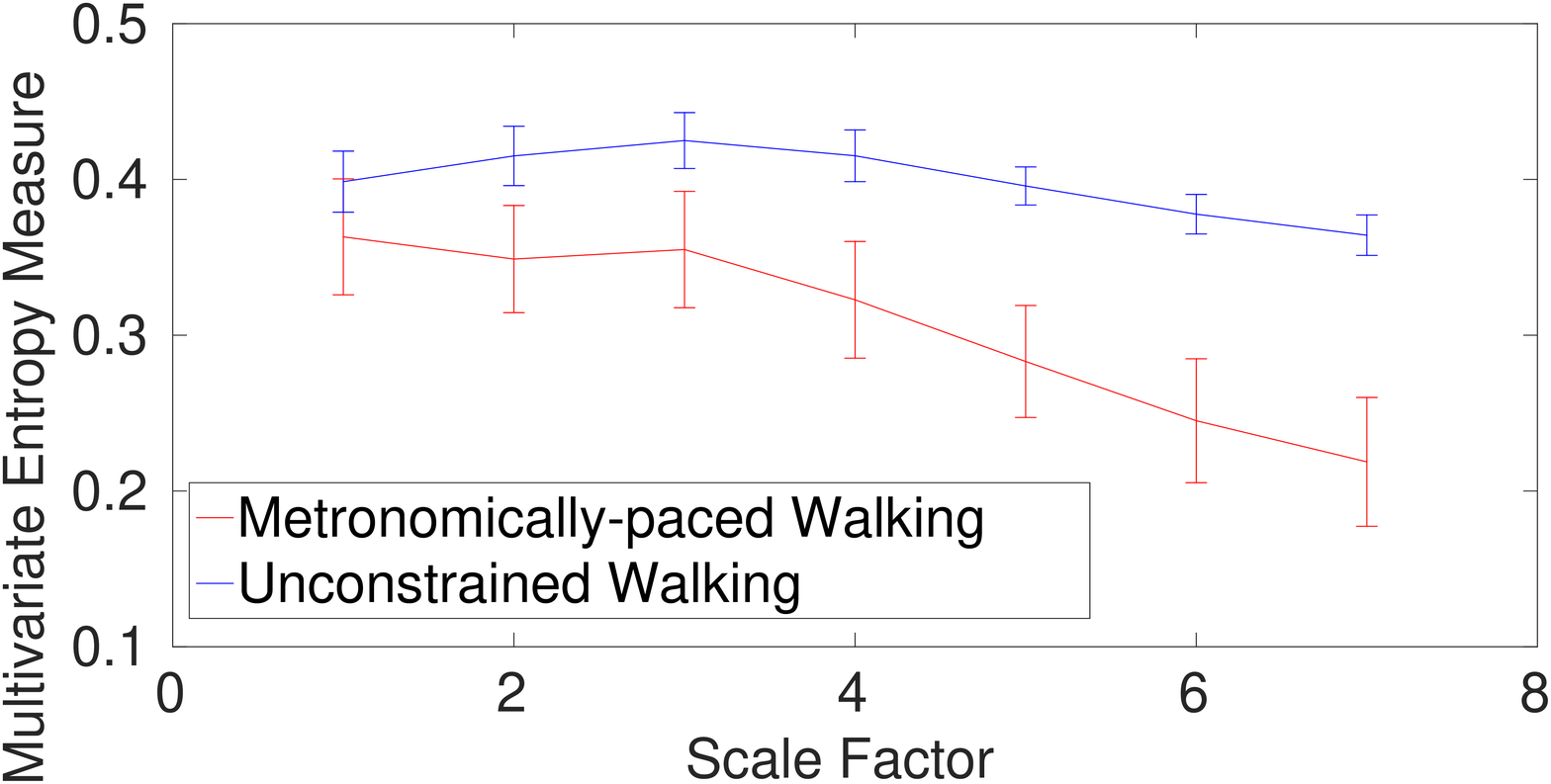}
		\footnotesize{(c) mvMFE}
	\end{multicols}
	
	\caption{Mean value and SD of the results using (a) mvMDE-III, (b) mvMDE, and (c) mvMFE for self-paced vs. metronomically-paced stride interval fluctuations.}
	\label{figurelabel}

\end{figure*}

\begin{table}
  
	\centering
	\label{table_example}
		\setlength{\tabcolsep}{3.5pt}
	\centering
	\caption{\footnotesize CV values of the entropy results at scale factor 4 using mvMDE-III, mvMDE, and mvMFE for self-paced walk (SPW) vs. metronomically-paced walk (MPW).}
	\centering     
	\begin{tabular}{c*{6}{c}}
		\setlength{\tabcolsep}{1pt}
		mvMFE& mvMFE&mvMDE-III& mvMDE-III&mvMDE&  mvMDE\\
		for SPW&for MPW&for SPW &for MPW&for SPW&for MPW\\
		\hline
		\setlength{\tabcolsep}{2pt}
		0.040 & 0.116 & 0.005  & 0.025& 0.002 & 0.019 \\
		\setlength{\tabcolsep}{2pt}
		
	\end{tabular}
	\setlength{\tabcolsep}{4pt}
\end{table}

\textit{2) Dataset of Focal and Non-focal Brain Activity}: For the focal and non-focal EEG recordings, the results obtained by mvMDE-II, mvMDE-III, mvMDE, and mvMFE, respectively depicted in Fig. 5 (a), (b), (c), and (d), show that the focal time series are less complex than the non-focal ones, in agreement with previous studies \cite{andrzejak2012nonrandomness}\cite{sharma2015application}. The CV values for the focal- and non-focal-based results at scale 6 are shown in Table V. All the mvMDE-based CV values are smaller than those using mvMFE, showing more stability of the results obtained by the proposed methods. Moreover, the CV values for mvMDE are smaller than those for mvMDE-III, and the latter ones are smaller than those for mvMDE-II, suggesting that the mvMDE leads to more stable profiles.

\begin{figure*}
	\centering

	\begin{multicols}{4}
		\centering
		\includegraphics[width=4.6cm,height=3.5cm]{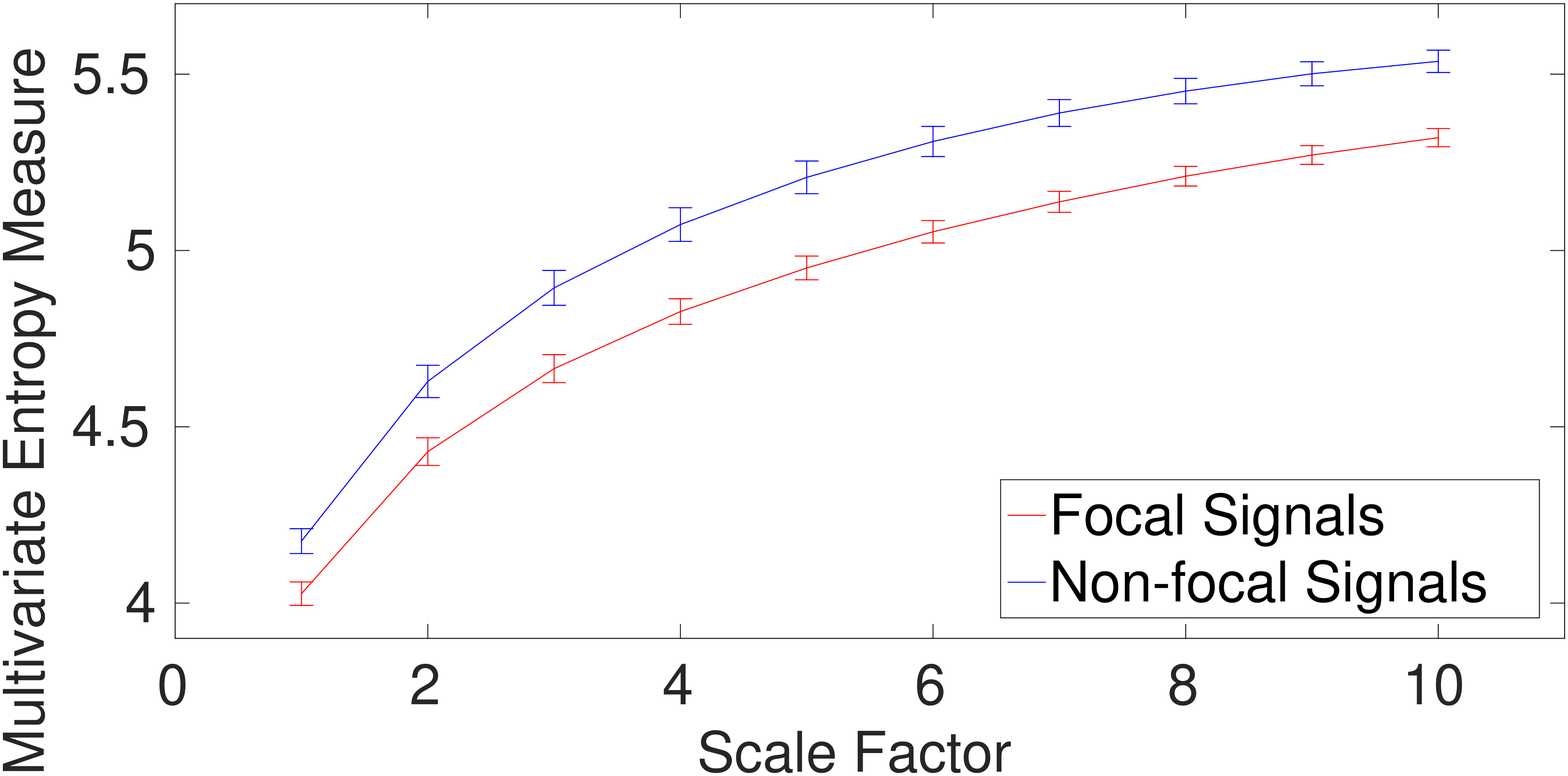}
		\footnotesize{(a) mvMDE-II}
		\includegraphics[width=4.6cm,height=3.5cm]{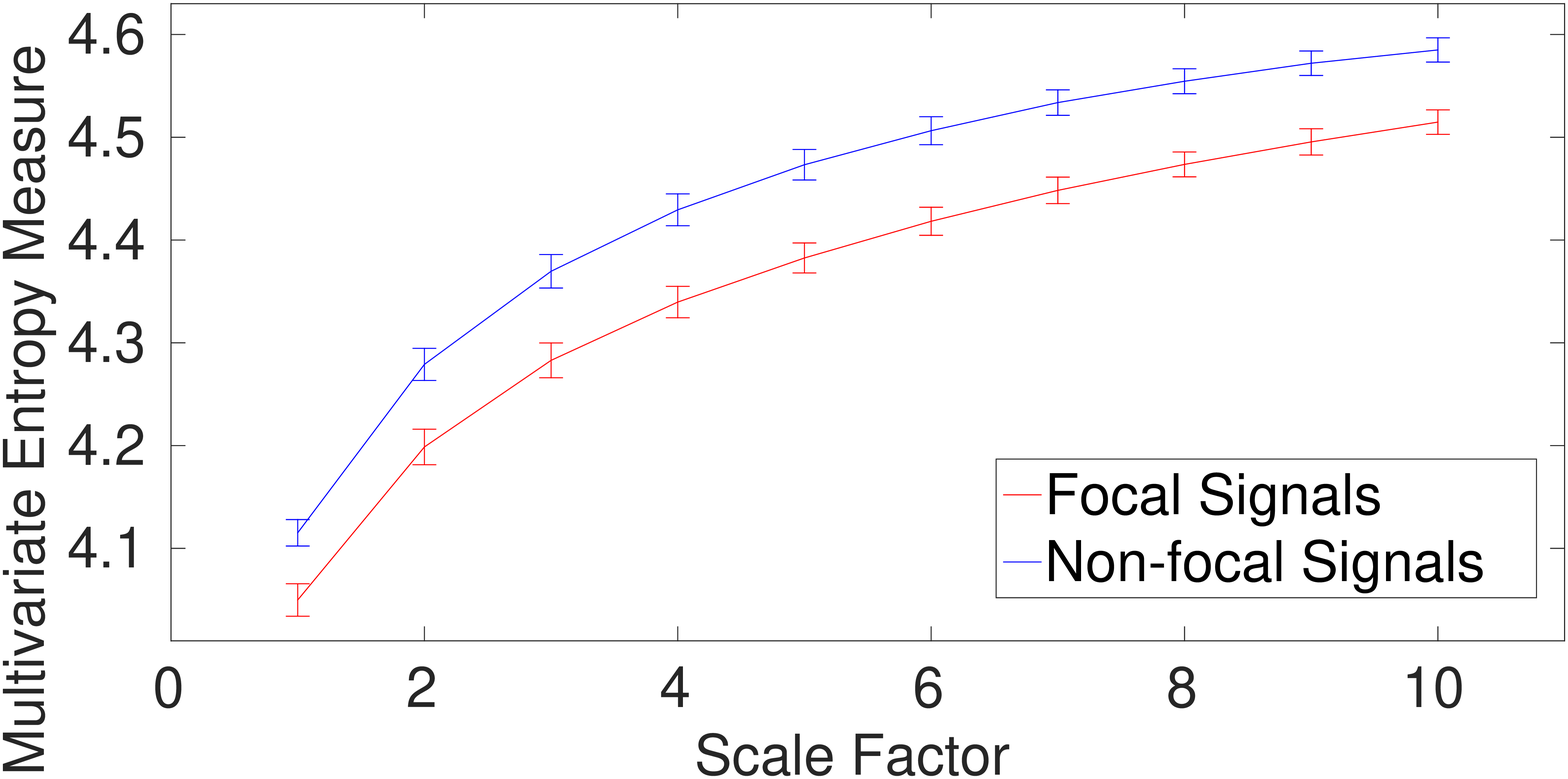}
		\footnotesize{(b) mvMDE-III}
		\includegraphics[width=4.6cm,height=3.5cm]{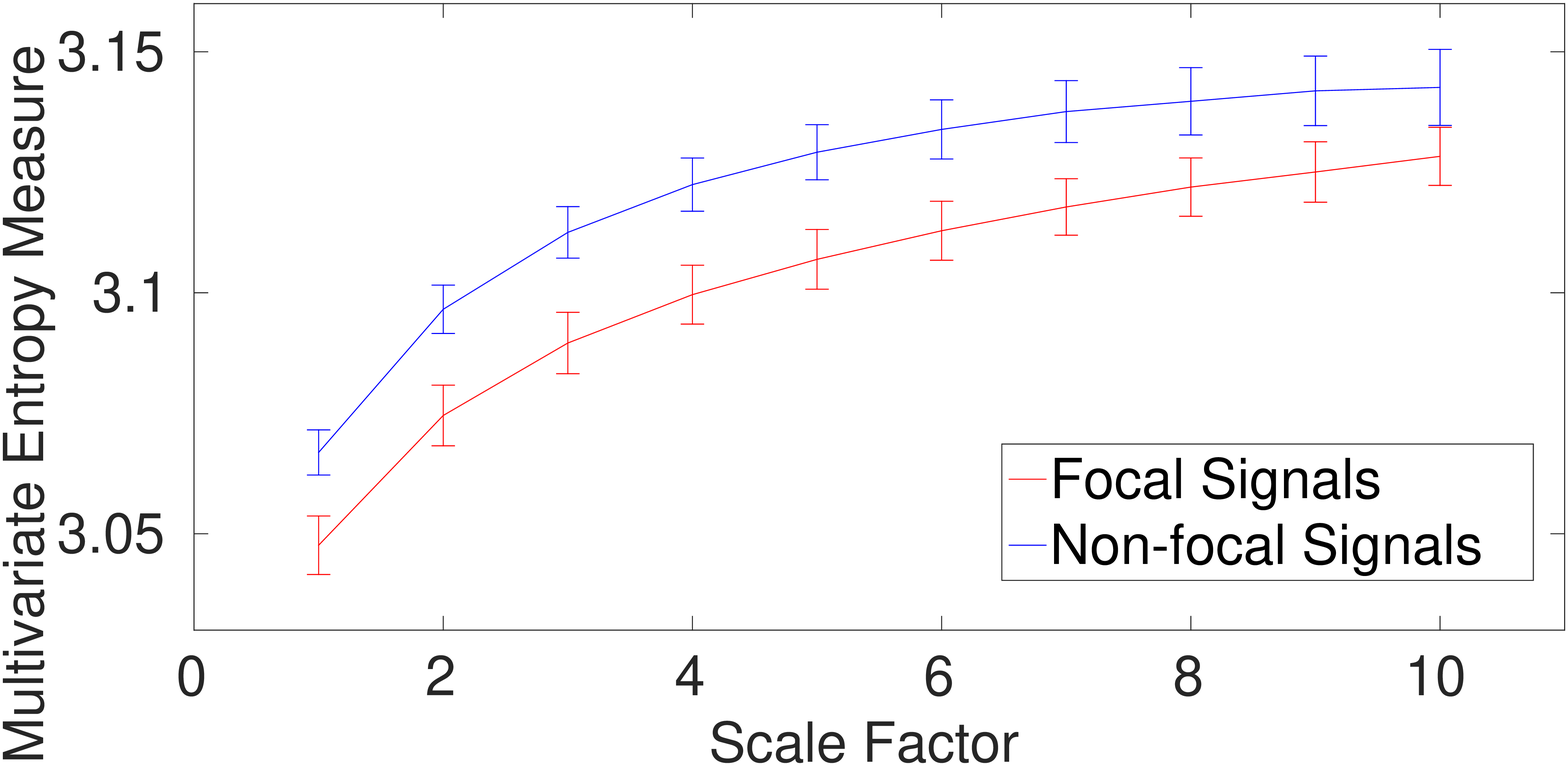}
		\footnotesize{(c) mvMDE}
		\includegraphics[width=4.6cm,height=3.5cm]{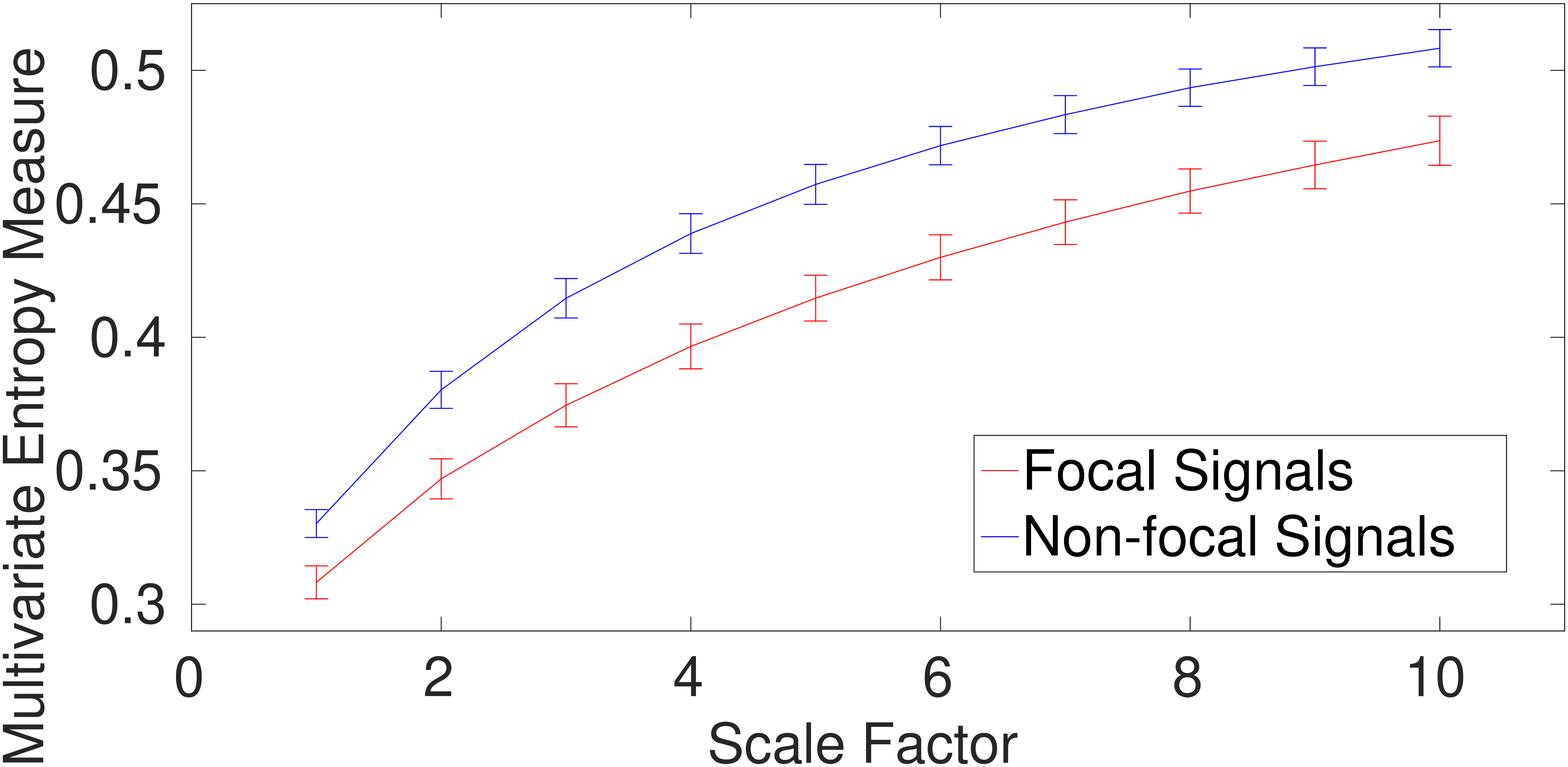}
		\footnotesize{(d) mvMFE}
		
	\end{multicols}
	
	\caption{Mean value and SD of the results using (a) mvMDE-II, (b) mvMDE-III, (c) mvMDE, and (d) mvMFE for focal vs. non-focal time series.}
	\label{figurelabel}

\end{figure*}

\begin{table}
	\setlength{\tabcolsep}{3pt}
	\centering
	\label{table_example}
	\centering
	\caption{\footnotesize CV values of the entropy results at scale factor 6 using mvMDE-II, mvMDE-III, mvMDE, and mvMFE for focal and non-focal EEG recordings.}
	\centering     
	\begin{tabular}{c*{8}{c}}
		\setlength{\tabcolsep}{1pt}
		mvMFE& mvMFE&mvMDE-II& mvMDE-II\\
		of focal signals&of non-focal signals&of focal signals&of non-focal signals\\
		\hline
		\setlength{\tabcolsep}{2pt}
		0.019 & 0.015 & 0.006  & 0.008\\
				\hline
		\setlength{\tabcolsep}{2pt}

		mvMDE-III&  mvMDE-III&mvMDE&  mvMDE\\
		of focal signals&of non-focal signals&of focal signals&of non-focal signals \\
		\hline
		\setlength{\tabcolsep}{2pt}
		0.003 & 0.003 & 0.002 & 0.002\\
		\setlength{\tabcolsep}{2pt}

	\end{tabular}
	\setlength{\tabcolsep}{2pt}
\end{table}

\textit{3) Surface MEG Recordings in Alzheimer's Disease}:
To assess the ability of mvMDE, in comparison with mvMFE, we applied the methods to the 148-channel MEG signals to discriminate AD patients from controls. Because mvMFE needs to store a huge number of elements for a signal with a large number of channels, mvMFE was not able to simultaneously analyse all 148 time series. However, the results using mvMDE are depicted in Fig. 6. It represents an advantage of mvMDE over mvMFE for signals with a large number of channels. To compare the mvMFE and mvMDE, we applied the methods to five main scalp regions, namely, anterior (17 channels), right (34 channels) and left lateral (34 channels), central (29 channels), and posterior (34 channels) areas, leading to the smaller number of channels to noticeably decrease the number of elements stored by the use of the mvMFE algorithm. 

The average and SD of mvMDE and mvMFE values for five regions are respectively shown in Fig. 7(a) and 7(b). As can be seen in Fig. 6 and Fig. 7, the average mvMDE and mvMFE values for AD patients are smaller than those for controls at lower scale factors (short-time scale factors), while at higher scales, the AD subjects' recordings have larger entropy values (long-time scale factors) for both the mvMFE and mvMDE, in agreement with \cite{yang2013cognitive,hornero2009nonlinear,azami2017univariate}. Because the larger the number of channels, the smaller the mvMSE and similarly mvMFE values \cite{azami2017univariate}, the entropy values for anterior region are larger than those for the other four regions. It is worth noting that we only use mvMDE, as the signals do not follow the minimum number of samples required for mvMDE-II and III.

The Mann–Whitney \textit{U}-test was used to assess the differences between the mvMDE and mvMFE profiles at each temporal scale for AD patients versus controls, because the mvMDE and mvMFE values at each scale factor did not follow a normal distribution. The temporal scales with \textit{p}-values smaller than 0.001 are shown with * in Fig. 6 and Fig. 7. The \textit{p}-values show that the mvMDE, compared with the mvMFE, significantly discriminated the controls from subjects with AD at a larger number of scale factors. Moreover, the smallest \textit{p}-value was achieved by the mvMDE, evidencing the superiority of mvMDE over mvMFE.

\begin{figure}
	\centering
	\includegraphics[width=7cm,height=3cm]{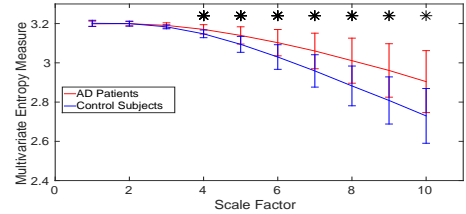}
	\caption{\small{Mean value and SD of the results obtained by mvMDE computed from 36 AD patients versus 26 elderly controls for all the 148 channels. Red and blue respectively indicate AD patients and controls. The scales with \textit{p}-values smaller than 0.001 are shown with *.}}
	\label{figurelabel}
\end{figure}

\begin{figure*}
	\centering
	\includegraphics[width=19cm,height=6cm]{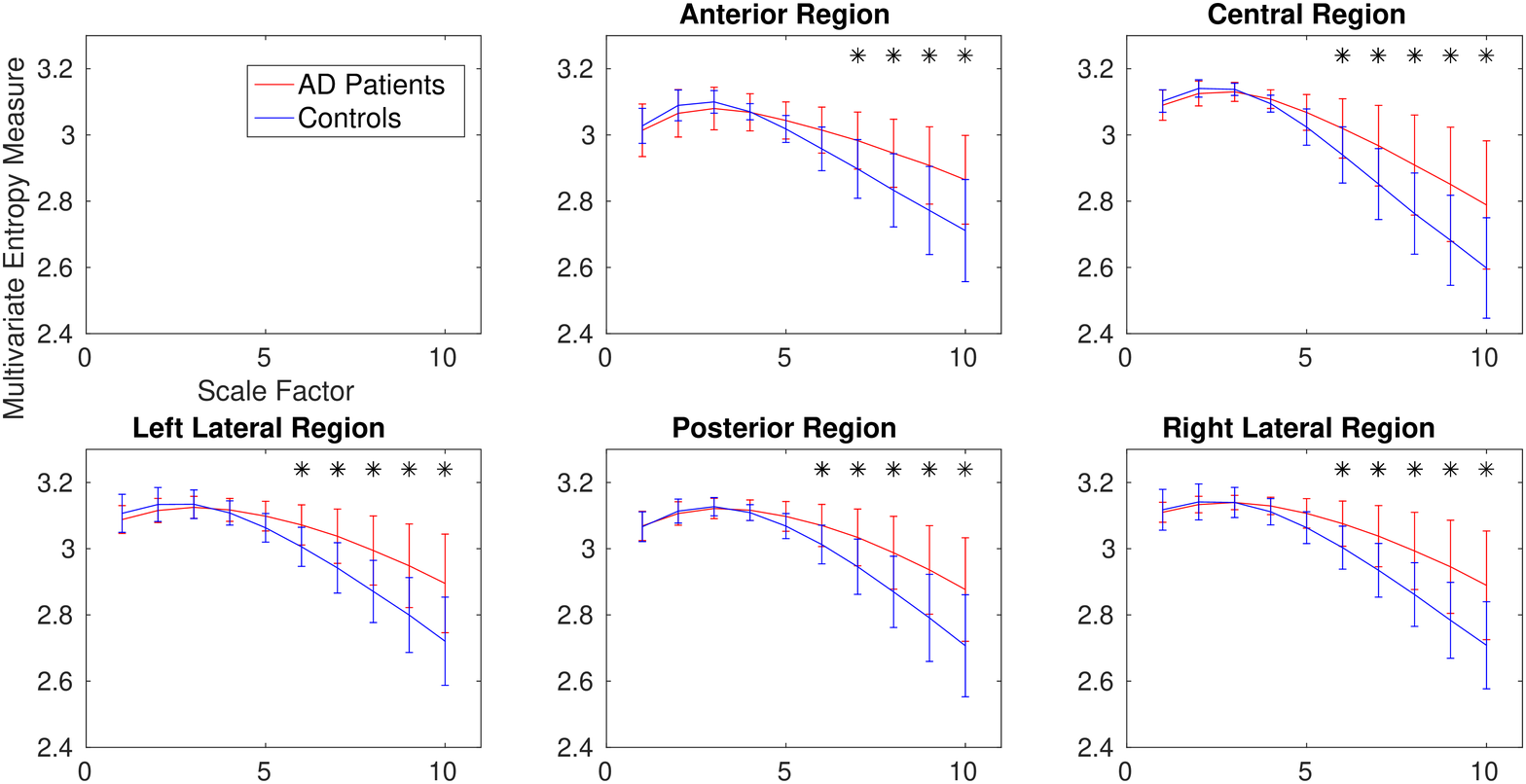}
	\small{(a) mvMDE}
	\includegraphics[width=19cm,height=6cm]{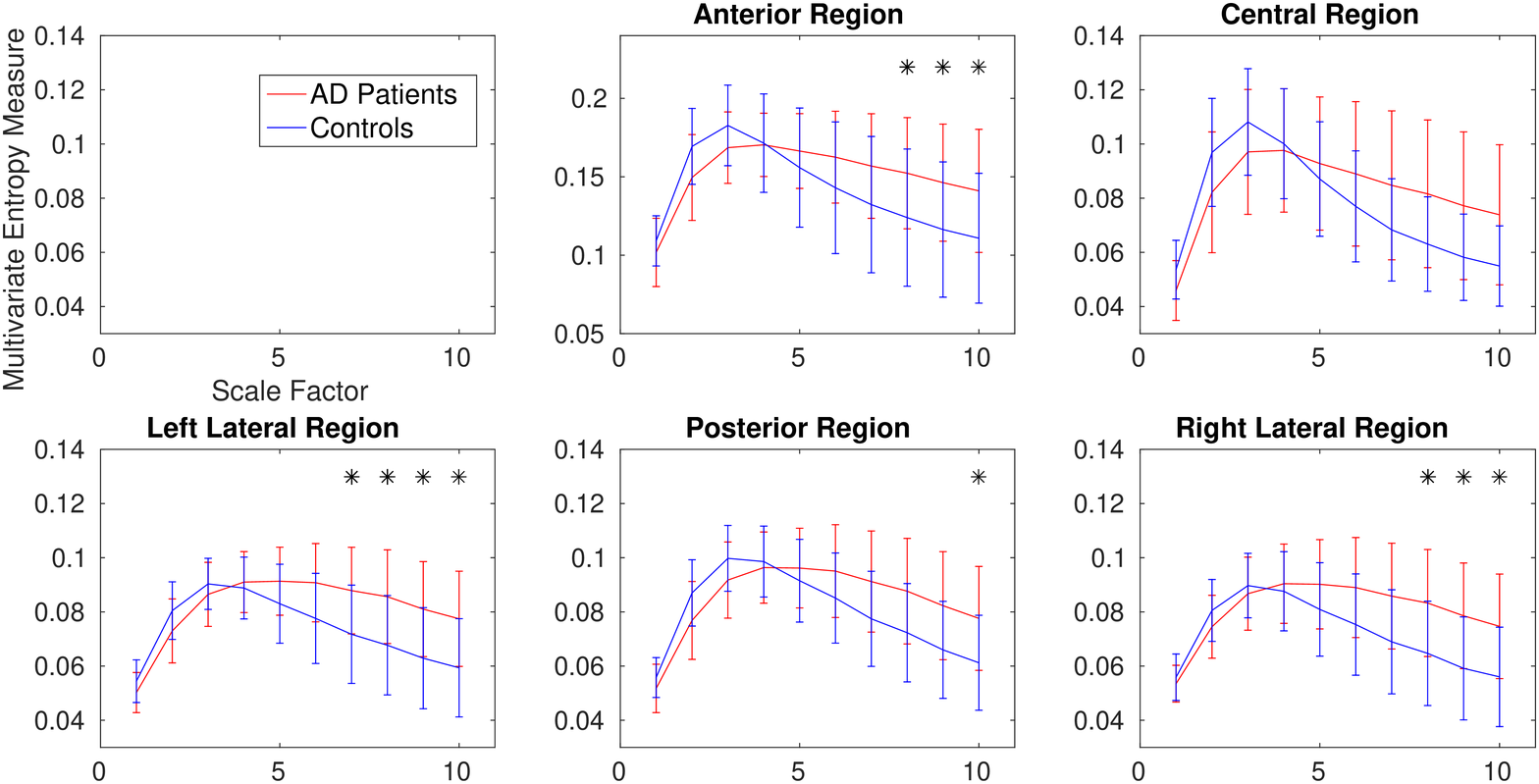}
	\small{(b) mvMFE}
	\\
	\caption{ Mean value and SD of the results obtained by (a) mvMDE and (b) mvMFE computed from 36 AD patients versus 26 elderly age-matched controls over five scalp regions. Red and blue indicate AD patients and controls, respectively. The scale factors with \textit{p}-values smaller than 0.001 are shown with *.}
	\label{figurelabel}
\end{figure*}

On the whole, the profiles for the real datasets evidence the advantage of mvMDE-II, mvMDE-III, and mvMDE over mvMFE to discriminate different types of dynamics of multi-channel signals as well as the superiority of mvMDE over mvMFE in terms of ability to discriminate various dynamics of time series, computational time, and memory cost. As mentioned before, mvMPE does not consider the spatial domain. We have also refined the mvMPE \cite{morabito2012multivariate} on the basis of mvMDE-II, mvMDE-III, and mvMDE. These approaches have the following advantages over the first version of mvMPE \cite{morabito2012multivariate}: 1) they take into account both the spatial and time domains; 2) their results were more stable than the mvMPE-based ones; and 3) better distinguished different dynamics of multivariate signals. However, since the mvMDE methods are considerably faster, result in more stable profiles, and lead to larger differences between physiological conditions of recordings, for simplicity, we did not report the mvMPE-based results. Our future study will aim at proposing the refined composite mvMDE (RCmvMDE) approaches according to \cite{azami2017refined}. Moreover, we will explore the mvMDE and RCmvMDE on other physiological and non-physiological time series.

\section{Conclusions}
To quantify the complexity of biomedical multivariate time series, we built four diverse alternative implementations of mvMDE as further developments of our recently introduced MDE \cite{azami2017refineddd}. These insights help towards a comprehensive understanding of four strategies to extend a univariate-based entropy method to its multivariate versions and therefore, provide invaluable information for future studies on multivariate time series. Although mvMDE was the best algorithm in terms of ability to discriminate dynamics of multivariate signals, computational time, and memory cost, the simpler alternatives (mvDE-I to III) may still be useful in some settings.

We assessed their performance on the correlated and uncorrelated multivariate noise signals, the bivariate AR time series, and three physiological datasets. The results showed the similar behavior of mvMSE-, mvMFE-, and mvMDE-based profiles. However, mvMDE had the following advantages over the existing methods: 1) it was noticeably faster than the existing methods; 2) mvMDE, in comparison with mvMSE and mvMFE, resulted in more stable profiles; 3) mvMDE better discriminated different kinds of biomedical signals; 4) for short multivariate time series, mvMDE, unlike mvMSE, did not result in undefined values; and 5) mvMDE, compared with mvMSE and mvMFE, needed to store a considerably smaller number of elements.

Overall, we expect the mvMDE approach to play a key role in the assessment of complexity in multivariate time series.

\section*{Acknowledgement}
The MATLAB codes of the mvMDE techniques and the refined versions of mvMPE will be made publicly-available upon publication.

\ifCLASSOPTIONcaptionsoff
  \newpage
\fi
{
	\bibliographystyle{ieeetr}
	\bibliography{REFmvMDE}

\begin{thebibliography}{10}

\bibitem{cerutti2009multiscale}
S.~Cerutti, D.~Hoyer, and A.~Voss, ``Multiscale, multiorgan and multivariate
  complexity analyses of cardiovascular regulation,'' {\em Philosophical
  Transactions of the Royal Society of London A: Mathematical, Physical and
  Engineering Sciences}, vol.~367, no.~1892, pp.~1337--1358, 2009.

\bibitem{cerutti2012multivariate}
S.~Cerutti, ``Multivariate and multiscale analysis of biomedical signals:
  Towards a comprehensive approach to medical diagnosis,'' in {\em
  Computer-Based Medical Systems (CBMS), 2012 25th International Symposium on},
  pp.~1--5, IEEE, 2012.

\bibitem{ahmed2012multivariate}
M.~U. Ahmed and D.~P. Mandic, ``Multivariate multiscale entropy analysis,''
  {\em IEEE Signal Processing Letters}, vol.~19, no.~2, pp.~91--94, 2012.

\bibitem{pereda2005nonlinear}
E.~Pereda, R.~Q. Quiroga, and J.~Bhattacharya, ``Nonlinear multivariate
  analysis of neurophysiological signals,'' {\em Progress in neurobiology},
  vol.~77, no.~1, pp.~1--37, 2005.

\bibitem{ahmed2011multivariate}
M.~U. Ahmed and D.~P. Mandic, ``Multivariate multiscale entropy: A tool for
  complexity analysis of multichannel data,'' {\em Physical Review E}, vol.~84,
  no.~6, p.~061918, 2011.

\bibitem{costa2005multiscale}
M.~Costa, A.~L. Goldberger, and C.-K. Peng, ``Multiscale entropy analysis of
  biological signals,'' {\em Physical review E}, vol.~71, no.~2, p.~021906,
  2005.

\bibitem{richman2000physiological}
J.~S. Richman and J.~R. Moorman, ``Physiological time-series analysis using
  approximate entropy and sample entropy,'' {\em American Journal of
  Physiology-Heart and Circulatory Physiology}, vol.~278, no.~6,
  pp.~H2039--H2049, 2000.

\bibitem{bandt2002permutation}
C.~Bandt and B.~Pompe, ``Permutation entropy: a natural complexity measure for
  time series,'' {\em Physical review letters}, vol.~88, no.~17, p.~174102,
  2002.

\bibitem{rostaghi2016dispersion}
M.~Rostaghi and H.~Azami, ``Dispersion entropy: A measure for time series
  analysis,'' {\em IEEE Signal Processing Letters}, vol.~23, no.~5,
  pp.~610--614, 2016.

\bibitem{fogedby1992phase}
H.~C. Fogedby, ``On the phase space approach to complexity,'' {\em Journal of
  statistical physics}, vol.~69, no.~1-2, pp.~411--425, 1992.

\bibitem{silva2015multiscale}
L.~E.~V. Silva, B.~C.~T. Cabella, U.~P. da~Costa~Neves, and L.~O.~M. Junior,
  ``Multiscale entropy-based methods for heart rate variability complexity
  analysis,'' {\em Physica A: Statistical Mechanics and its Applications},
  vol.~422, pp.~143--152, 2015.

\bibitem{azami2017refined}
H.~Azami and J.~Escudero, ``Refined composite multivariate generalized
  multiscale fuzzy entropy: A tool for complexity analysis of multichannel
  signals,'' {\em Physica A: Statistical Mechanics and its Applications},
  vol.~465, pp.~261--276, 2017.

\bibitem{li2014coupling}
P.~Li, L.~Ji, C.~Yan, K.~Li, C.~Liu, and C.~Liu, ``Coupling between short-term
  heart rate and diastolic period is reduced in heart failure patients as
  indicated by multivariate entropy analysis,'' in {\em Computing in Cardiology
  Conference (CinC), 2014}, pp.~97--100, IEEE, 2014.

\bibitem{morabito2012multivariate}
F.~C. Morabito, D.~Labate, F.~La~Foresta, A.~Bramanti, G.~Morabito, and
  I.~Palamara, ``Multivariate multi-scale permutation entropy for complexity
  analysis of {Alzheimer's} disease {EEG},'' {\em Entropy}, vol.~14, no.~7,
  pp.~1186--1202, 2012.

\bibitem{yin2017multivariate}
Y.~Yin and P.~Shang, ``Multivariate weighted multiscale permutation entropy for
  complex time series,'' {\em Nonlinear Dynamics}, pp.~1--16, 2017.

\bibitem{azami2017univariate}
H.~Azami, D.~Ab{\'a}solo, S.~Simons, and J.~Escudero, ``Univariate and
  multivariate generalized multiscale entropy to characterise {EEG} signals in
  {Alzheimer's} disease,'' {\em Entropy}, vol.~19, no.~1, p.~31, 2017.

\bibitem{labate2013entropic}
D.~Labate, F.~La~Foresta, G.~Morabito, I.~Palamara, and F.~C. Morabito,
  ``Entropic measures of eeg complexity in alzheimer's disease through a
  multivariate multiscale approach,'' {\em Sensors Journal, IEEE}, vol.~13,
  no.~9, pp.~3284--3292, 2013.

\bibitem{zhao2016multivariable}
L.~Zhao, S.~Wei, H.~Tang, and C.~Liu, ``Multivariable fuzzy measure entropy
  analysis for heart rate variability and heart sound amplitude variability,''
  {\em Entropy}, vol.~18, no.~12, p.~430, 2016.

\bibitem{ramdani2016parameters}
S.~Ramdani, V.~Bonnet, G.~Tallon, J.~Lagarde, P.~L. Bernard, and H.~Blain,
  ``Parameters selection for bivariate multiscale entropy analysis of postural
  fluctuations in fallers and non-fallers older adults,'' {\em IEEE
  Transactions on Neural Systems and Rehabilitation Engineering}, vol.~24,
  no.~8, pp.~859--871, 2016.

\bibitem{azami2017refineddd}
H.~Azami, M.~Rostaghi, D.~Abasolo, and J.~Escudero, ``Refined composite
  multiscale dispersion entropy and its application to biomedical signals,''
  {\em IEEE Transactions on Biomedical Engineering}, DOI:
  10.1109/TBME.2017.2679136, 2017.

\bibitem{cao1998dynamics}
L.~Cao, A.~Mees, and K.~Judd, ``Dynamics from multivariate time series,'' {\em
  Physica D: Nonlinear Phenomena}, vol.~121, no.~1, pp.~75--88, 1998.

\bibitem{humeau2016multivariate}
A.~Humeau-Heurtier, ``Multivariate generalized multiscale entropy analysis,''
  {\em Entropy}, vol.~18, no.~11, p.~411, 2016.

\bibitem{penny2002bayesian}
W.~Penny and S.~Roberts, ``Bayesian multivariate autoregressive models with
  structured priors,'' {\em IEE Proceedings-Vision, Image and Signal
  Processing}, vol.~149, no.~1, pp.~33--41, 2002.

\bibitem{ahmed2012dynamical}
M.~Ahmed, N.~Rehman, D.~Looney, T.~Rutkowski, and D.~Mandic, ``Dynamical
  complexity of human responses: a multivariate data-adaptive framework,'' {\em
  Bulletin of the Polish Academy of Sciences: Technical Sciences}, vol.~60,
  no.~3, pp.~433--445, 2012.

\bibitem{hausdorff1996fractal}
J.~M. Hausdorff, P.~L. Purdon, C.~Peng, Z.~Ladin, J.~Y. Wei, and A.~L.
  Goldberger, ``Fractal dynamics of human gait: stability of long-range
  correlations in stride interval fluctuations,'' {\em Journal of applied
  physiology}, vol.~80, no.~5, pp.~1448--1457, 1996.

\bibitem{andrzejak2012nonrandomness}
R.~G. Andrzejak, K.~Schindler, and C.~Rummel, ``Nonrandomness, nonlinear
  dependence, and nonstationarity of electroencephalographic recordings from
  epilepsy patients,'' {\em Physical Review E}, vol.~86, no.~4, p.~046206,
  2012.

\bibitem{escudero2011regional}
J.~Escudero, S.~Sanei, D.~Jarchi, D.~Ab{\'a}solo, and R.~Hornero, ``Regional
  coherence evaluation in mild cognitive impairment and {Alzheimer's} disease
  based on adaptively extracted magnetoencephalogram rhythms,'' {\em
  Physiological measurement}, vol.~32, no.~8, p.~1163, 2011.

\bibitem{Labate2013}
D.~Labate, F.~La~Foresta, G.~Morabito, I.~Palamara, and F.~C. Morabito,
  ``Entropic measures of {EEG} complexity in {Alzheimer's} disease through a
  multivariate multiscale approach,'' {\em IEEE Sensors Journal}, vol.~13,
  no.~9, pp.~3284--3292, 2013.

\bibitem{sharma2015application}
R.~Sharma, R.~B. Pachori, and U.~R. Acharya, ``Application of entropy measures
  on intrinsic mode functions for the automated identification of focal
  electroencephalogram signals,'' {\em Entropy}, vol.~17, no.~2, pp.~669--691,
  2015.

\bibitem{yang2013cognitive}
A.~C. Yang, S.-J. Wang, K.-L. Lai, C.-F. Tsai, C.-H. Yang, J.-P. Hwang, M.-T.
  Lo, N.~E. Huang, C.-K. Peng, and J.-L. Fuh, ``Cognitive and neuropsychiatric
  correlates of {EEG} dynamic complexity in patients with {Alzheimer's}
  disease,'' {\em Progress in Neuro-Psychopharmacology and Biological
  Psychiatry}, vol.~47, pp.~52--61, 2013.

\bibitem{hornero2009nonlinear}
R.~Hornero, D.~Ab{\'a}solo, J.~Escudero, and C.~G{\'o}mez, ``Nonlinear analysis
  of electroencephalogram and magnetoencephalogram recordings in patients with
  {Alzheimer's} disease,'' {\em Philosophical Transactions of the Royal Society
  of London A: Mathematical, Physical and Engineering Sciences}, vol.~367,
  no.~1887, pp.~317--336, 2009.

\end{thebibliography}
}

\end{document}